\documentclass[article,12pt,twocolumn]{article}

\usepackage[utf8]{inputenc}
\usepackage[russian,english]{babel}
\usepackage[T1]{fontenc}
\usepackage{hyperref}

\usepackage{ucs}
\usepackage{latexsym}
\usepackage{float}
\usepackage{mathtools}
\usepackage{amssymb}
\usepackage{amstext}
\usepackage{amsfonts}
\usepackage{graphicx}
\usepackage{ragged2e}
\usepackage{xcolor}
\usepackage{footmisc}
\usepackage{commath}

\usepackage{helvet}
\usepackage{textcomp}
\usepackage{mathabx}
\usepackage{multicol}

\usepackage[font=footnotesize,labelfont=bf]{caption}
\addto\captionsrussian{}

\usepackage{geometry}
\begin{document}
\selectlanguage{english}
\onecolumn

\center{\textbf{\Large{Dynamic model of a non-equilibrium chemical
composition formation in the shell of single neutron stars}}}

\textbf{\normalsize{A.Yu.Ignatovskiy$^{\text{\textbf{1,2,3,4,}}\asterisk}$, G.S.Bisnovatyi-Kogan$^{\text{\textbf{1,2,5}}}$}}

\footnotesize{$^{\text{\textsl{1}}}$\textsl{Space research institute RAS}\\
$^{\text{\textsl{2}}}$\textsl{Moscow institute of physics and technology}\\
$^{\text{\textsl{3}}}$\textsl{Institute for theoretical and experimental physics, NRC "Kurchatov Institute"}\\
$^{\text{\textsl{4}}}$\textsl{National research center "Kurchatov Institute"}\\
$^{\text{\textsl{5}}}$\textsl{Moscow engineering physics institute}\\
$^{\text{\textsl{*}}}$\textsl{E-mail: Lirts@phystech.edu}}

\justifying
\vspace{0.5cm}
\noindent
\small
\textbf{Abstract}--The process of a non-equilibrium chemical composition formation in the crust of a new born neutron star is considered, during cooling due to neutrino energy losses. A model is constructed for explaining accumulation of a large quantity of nuclear energy, which can maintain the X-ray luminosity of such compact objects for a long period of time. We studied numerically the dependence of the final chemical composition on various parameters of the model.

\vspace{0.5cm}
\noindent
\normalsize

\begin{multicols}{2}
\center{1. INTRODUCTION} \\ \justifying Neutron stars are the result of a massive star gravitational collapse at the end of its evolution. Densities of neutron stars are of the order of the atomic nucleus density $ \sim 2 \cdot 10^{14}$ g $\cdot$ $\mathrm{cm}^{- 3} $ in the inner parts and are $\sim 10^{9} $ g $\cdot$ $\mathrm{cm}^{-3}$ on the outer shell boundary. Newly formed neutron stars with inner temperatures $ \mathrm{T_{9}\gg 10} $ (here and below $\mathrm{T_{9} = T / 10^{9} \, K} $) cool rapidly, in first seconds of life, due to neutrino energy losses.

The results of the chemical composition evolution in various systems are of interest to all astrophysics. The answer is well known for the slow stellar evolution, when, as a result of various nuclear reactions, elements of the iron group are formed, which have the highest binding energy per nucleon $ \mathrm {\Delta E_{56Fe} \approx 8.55} $ MeV, and up to heavier elements such as platinum and plumbum. The processes of nucleosynthesis were classified in the work \cite{B2FH}, where authors figured out that elements heavier than iron are formed in the so-called s- and r-processes, the processes of slow and rapid capture of neutrons by nuclei respectively, followed by beta decays \cite{cameron2,seeger}. For the s-process, the relation between the probabilities of a neutron capture and an electron decay satisfies the inequality $ \mathrm {\lambda_ {n \gamma} \ll \lambda_{\beta}} $, and its path runs along the stable isotopes. For the r-process, the reverse inequality holds $ \mathrm {\lambda_{n \gamma} \gg \lambda_{\beta}} $ \cite{hillebrandt, r-process, PanovJanka} and its path is far from the stability valley of the chemical elements. No experimental nuclear data on strongly neutron-rich nuclei are available.
For the nuclei production of the heaviest transuranic elements in the r-process, up to 150 free neutrons are required per seed nucleus with a mass number of $\mathrm{A = 50-100}$.
Astrophysical objects associated with the possible nucleosynthesis of extremely heavy elements are neutron stars. The formation of heavy elements in a neutron bunch was first considered in the work \cite{Mayer_Teller}. The conditions for the r-process to occur are realized during the evolution of a close binary system of neutron stars of different masses \cite{PanovYudin1}, during the merging of neutron stars \cite{FreiThiel}, and in a hot wind from the surface of neutron stars \cite{ArconesThielemann}.

The aim of the current work is to carry out numerical calculations of the chemical composition evolution in the shell of the hot, newly formed neutron star that rapidly cools due to neutrino energy losses. This problem was first formulated and discussed in the work \cite{B-kogan1}. The calculation for the relaxation of the nuclear composition at some fixed temperature and density values had been made earlier in the work \cite{Ignatovskiy}.

\center{2. SCHEME OF A NON-EQUILIBRIUM LAYER FORMATION} \\ \justifying A key role in the formation of the non-equilibrium layer belongs to a gradual and non-simultaneous closure of various nuclear reaction channels during cooling \cite{noneq layer}. First, a fusion of heavy charged nuclei and protons capture by nuclei are ceased because the probability of Coulomb barrier tunneling decreases with a temperature drop \cite{ll_q_mech}), and photonuclear reactions stop later.

Initially, when a matter is at high temperature and density, nuclei, protons and neutrons are non-degenerate $\mathrm{kT > \varepsilon_{fe}}$, non-relativistic $\mathrm{p_{fe} <  m_p c}$ and are described by Boltzmann gas. Under these conditions, the matter is in equilibrium relative to strong interaction reactions \cite{nse,waitingpoint}:
\begin{equation}
\label{NSE}
\mathrm{\mu(A,Z)+\mu_{i}=\mu(A',Z')}.
\end{equation}
Here we denote $\mathrm{i=n,p}$. Using the expression for the chemical potential $\mu$ of the Boltzmann gas, one can obtain the Saha equation \cite{ll_s_ph} that determines the concentration of each nucleus in the system as a function of free neutrons and protons concentrations in the following form:
\begin{equation}
\label{Saha}
\begin{aligned}
&\mathrm{n(A,Z)=g(A,Z)\frac{n_{p}^{Z}n_{n}^{A-Z}}{2^{A}} A^{\frac{3}{2}}\times}\\
&\mathrm{\times \bigg( \frac{2\pi \hbar^{2}}{m_{\mu}kT}\bigg)^{\frac{3}{2}(A-1)} \exp \bigg\{ \frac{Q(A,Z)}{kT} \bigg\}. }
\end{aligned}
\end{equation}
Here $\mathrm{Q(A,Z)=c^{2}[Zm_{p}+(A-Z)m_{n}-m(A,Z)]}$ is the nuclear binding energy.
Electrons are relativistic $\mathrm{p_{fe} \gg m_{e} c} $ \cite {ll_f_th} and degenerate $\mathrm{kT < \varepsilon_{fe}} $, the degeneracy degree increases with cooling $ \mathrm{kT \ll \varepsilon_{fe}}$. Cooling occurs mainly due to neutrino energy losses during the following reactions:
\begin{equation}
\label{beta-reactions}
\begin{cases}
  \mathrm{\; \; (A,Z)+e^{\pm}\rightarrow(A,Z\pm 1)+ \nu^{\pm}},\\
  \mathrm{\; \; (A,Z\pm 1) \rightarrow (A,Z) + e^{\pm} + \nu^{\mp}},
\end{cases}
\end{equation}
denoting $\mathrm{\nu^{\pm}}$ as antineutrino and neutrino, respectively. It is necessary to take into account positrons in equilibrium at $ \mathrm{kT \gtrsim m_{e} c^{2}}$, what corresponds to temperatures $ \mathrm{T_{9} \gtrsim 6}$. The condition for the annihilation reaction equilibrium of the pair $\mathrm{e^{-} + e^{+} \leftrightarrows 2 \gamma}$ for $\mu_\gamma=0 $ leads to the relation $\mathrm{\mu_{e^{-}}+\mu_{e^{+}}=0}$.

As a result of a temperature drop, the reactions with protons freeze out and positrons disappear. Only beta reactions with electrons as well as capture of neutrons and photonuclear reactions remain at relatively low temperatures. Stable nuclei rapidly capture free neutrons ($\mathrm{\tau_{n \gamma} \ll \tau_{\gamma n}}$ and $\mathrm{\tau_{n \gamma} \ll \tau_{\beta}} , \, \, \, \tau$ -- characteristic reaction time), moving away from the stability valley to the region of low binding energy values of neutrons in the nuclei, until direct reactions of neutron capture are balanced by photonuclear reactions ($\mathrm{\tau_{n \gamma} \sim \tau_{\gamma n}}$). As a result, formation of heavy nuclei occurs, with large atomic numbers $ \mathrm{A_{j}} $ for each isotopic chain $\mathrm{Z_{i}}$.

Electrons degenerate and occupy the low energy levels under the Fermi surface. The Pauli exclusion principle restrict beta reactions: the nucleus can either decay at $\mathrm{Q_{\beta} = c^2 [m(A,Z-1)-m(A, Z)-m_e]> \varepsilon_{fe}}$, or capture electrons at $\mathrm{\varepsilon_{fe}> Q_{\beta} = \varepsilon_{\beta i}}$, which is a threshold Fermi energy for electron capture. As a result of these processes, nuclei accumulate in a non-equilibrium layer, limited in $\mathrm{Z}$ and $\mathrm{A}$ \cite{B-kogan1}.
\begin{figure*}[ht!]
  \includegraphics[width=\textwidth,height=0.4\textheight]{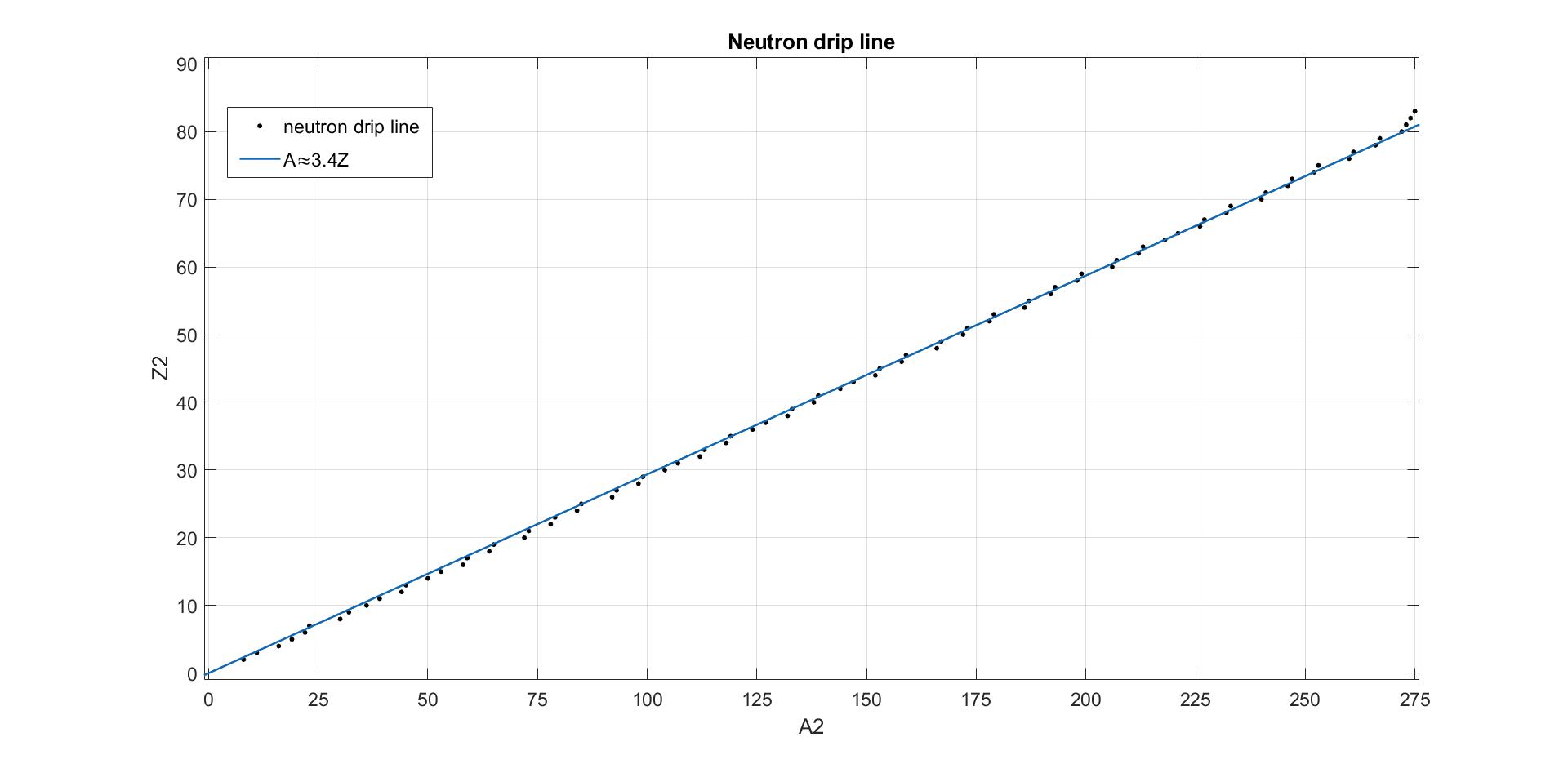}
  \caption{\footnotesize{Neutron drip line based on the finite range droplet model \cite{FRDM} for $\mathrm{Z \in [0;\,83]}$.}}
  \label{A(Z)}
\end{figure*}

In addition to the formation of a nuclear composition far from equilibrium \cite{bet1,bet2}, disequilibrium also consists in a large excess of free neutrons. Free neutrons can neither decay ($\mathrm{\varepsilon_{fe}>Q_{\beta n}=0.78}$ MeV), nor join to neutron-rich nuclei with low neutron binding energies. As a result, the matter does not reach the minimum energy state. The number of allowed nuclei decreases with a temperature drop \cite{B-kogan1}.

\begin{figure*}[hb!]
  \includegraphics[width=\textwidth,height=0.4\textheight]{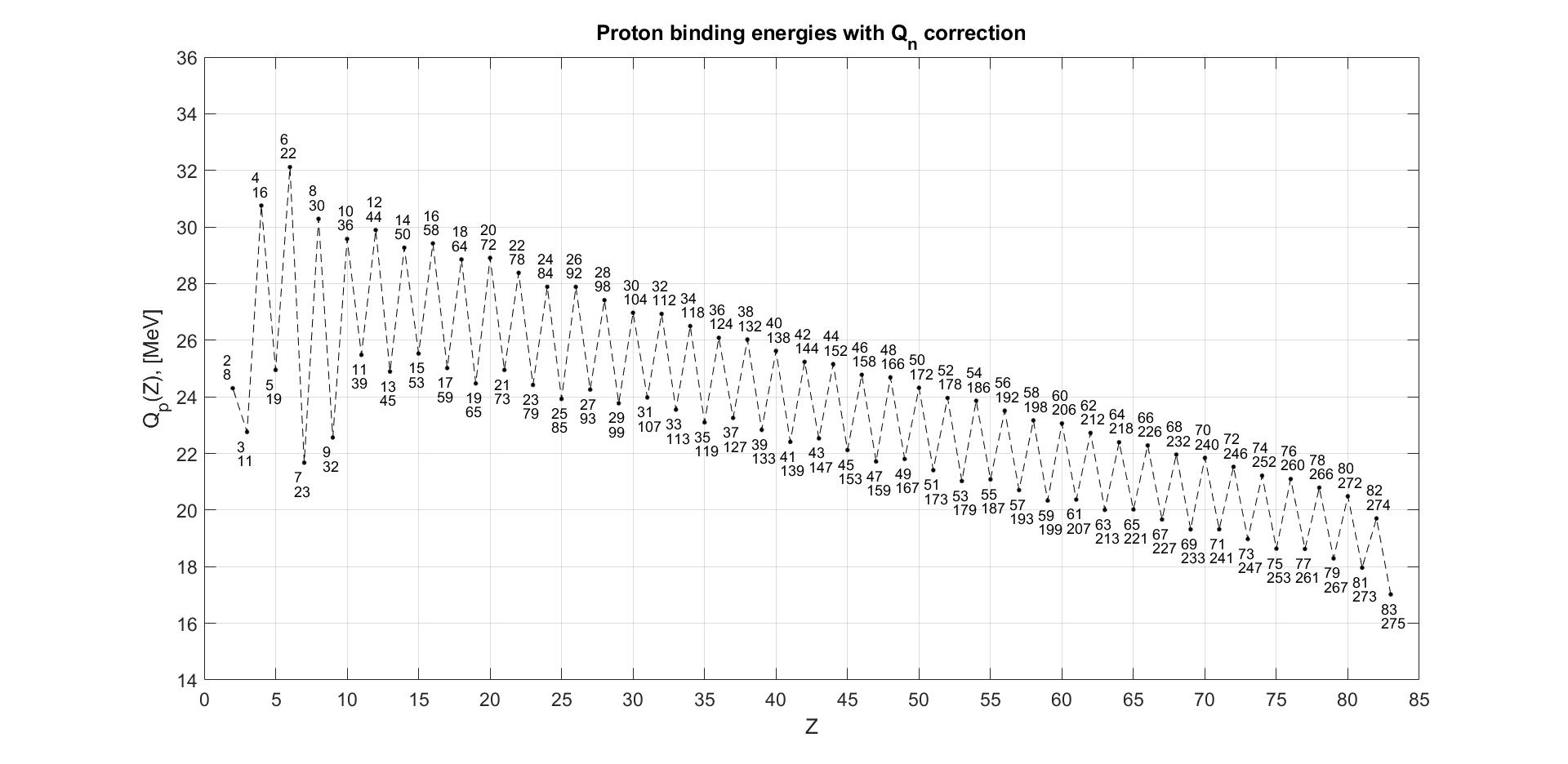}
  \caption{\footnotesize{Proton binding energies of nuclei located on the neutron drip line $\mathrm{Q_{n}=0}$ for $\mathrm{Z \in [0;\,83]}$. Numbers mark the atomic charge $\mathrm{Z}$ at the top and the atomic mass $\mathrm{A}$ at the bottom.}}
  \label{Qp(Z)_correction}
\end{figure*}

One can estimate the range of densities that makes a non-equilibrium layer formation possible. To evaluate the upper boundary, consider the evolution of matter with increasing density and at low temperature. For the threshold energy of electron capture, we approximately have $\mathrm{\varepsilon_{\beta i} \approx Q_{p}(A, Z) -Q_{n}(A, Z-1)}$, where $\mathrm{Q_{i}}$ is the binding energy of the particle \textsl{i} in the nucleus. As the density increases, the Fermi energy of electrons $\mathrm{\varepsilon_{fe} \approx \hbar c(3 \pi n_{e})^{1/3}}$ increases also. Upon reaching $\mathrm{\varepsilon_{fe} = \varepsilon_{\beta i}}$ nuclei capture electrons, decreasing their atomic charge numbers $\mathrm{Z}$, until they reach the neutron drip line $\mathrm{Q_{n} = 0}$, at which $\mathrm{\varepsilon_{\beta i}=Q_{p}-Q_{n}=Q_{p0}}$. The neutron drip line is illustrated in Fig.\ref{A(Z)}, based on the finite range droplet model \cite{FRDM} (see Section 3). At the moment the nucleus reaches $\mathrm{Q_ {n} = 0}$, subsequent electron captures will lead to the neutrons releasing from the nucleus \cite{B-kogan1}:
\begin{equation}
\label{3particles}
\mathrm{(A,Z)+e^{-} \rightarrow (A-1,Z-1)+n+\nu^{-}}.
\end{equation}
The described process repeats until the nucleus reaches the maximum $\mathrm{\varepsilon_{\beta max}=Q_{p0}^{(max)} \approx 32} $ MeV, which corresponds to carbon $\mathrm{^{22}C}$. Fig.\ref{Qp(Z)_correction} illustrates the proton binding energies in nuclei on the neutron drip line $\mathrm{Q_{n}=0}$. Proton binding energies in the used mass model are calculated as the following:
\begin{equation}
\label{Qp}
\begin{split}
\mathrm{Q_{p}(A,Z)}&=\mathrm{ Q(A,Z) - Q(A-1,Z-1)}\\
&\mathrm{\; - \abs{Q_{n}(A-1,Z-1)}}.
\end{split}
\end{equation}
The correction for the negative neutron binding energies $\mathrm{Q_{n}(A-1, Z-1) <0 \,}$ of the nuclei is taken into account. With further compression of the matter, beta reactions become non-equilibrium $\mathrm{\varepsilon_{fe}> Q_{p0}^{(max)}}$ that can lead to reheating up to a temperature of $\mathrm{T_{9} \sim 5}$, at which nuclear statistical equilibrium is established. The maximum density at which matter consists of nuclei located on the neutron drip line $\mathrm{Q_{n}=0}$ is:
\begin{equation}
\begin{aligned}
\label{rhomax}
  &\rho_{\max }=\mu_{\mathrm{z}} \cdot 10^{6}\left(\frac{\varepsilon_{\mathrm{fe}}}{\mathrm{m}_{\mathrm{e}} \mathrm{c}^{2}}\right)^{3}=\\
  &=\mu_{\mathrm{z}} \cdot 10^{6}\left(\frac{\mathrm{Q}_{\mathrm{p} 0}^{(\max )}}{\mathrm{m}_{\mathrm{e}} \mathrm{c}^{2}}\right)^{3} \sim 10^{12} \text{ g $\cdot$ cm$^{-3}$},
\end{aligned}
\end{equation}
where $\mathrm{\mu_{z}} \ge 4$ is the number of baryons per electron. The paper \cite{sato} demonstrates that when one take into account pycnonuclear fusion reactions with neutrons evaporation, the upper density limit of the non-equilibrium layer can reach $\rho_{\max,p} \sim 5 \cdot  10^{13} {\text{ g}} \cdot {\text{cm}}^{-3}$.

To evaluate the minimal density, one can proceed as the following \cite{B-kogan1}. For $\mathrm{Z>10}$, the proton binding energies $\mathrm{Q_{p0}}$ at the neutron drip line $\mathrm{Q_{n}=0}$ can be approximated by the linear function \cite{chechetkin}:
\begin{equation}
\label{5}
\mathrm{\mathrm{Q_{p0}=29-\frac{Z}{8}} \text{ MeV}},
\end{equation}
where the relation $\mathrm{A \approx 3.4Z} $ is satisfied. We find an approximate dependence of the chemical composition on density in the following form:
\begin{equation}
\label{6}
\mathrm{\mathrm{Z=8 \bigg[29-0.511\bigg(10^{-6}\rho/\mu_{z}\bigg)^{1/3} \bigg]}}.
\end{equation}
According to the equation (\ref{6}) it follows that both values $\mathrm{(A, Z)}$ grow with decreasing density. At high values of $\mathrm{(A,Z)}$ nuclei become unstable with respect to fission and alpha decay reactions. As a result of spontaneous fission, the total number of nuclei increases continuously until all free neutrons are absorbed by them.
According to the paper \cite{chechetkin} for nuclei at the neutron drip line $\mathrm{Q_{n}=0}$, the evaluated half-life is:
\begin{equation}
\label{7}
\mathrm{\log T_{1/2}=157-0.93Z}.
\end{equation}
We find out that for $\mathrm{Z>153}$ the decay half-life is less than $\mathrm{10^{7}}$ years. Using the expression (\ref{5}) for $\mathrm{Z=153}$, we obtain that $\mathrm{Q_{p0}^{(min)} \approx 8}$ MeV. Similarly to the expression (\ref{rhomax}), we have the following estimation for the minimum density:
\begin{equation}
\label{8}
\mathrm{\rho_{min}=\mu_{z}\cdot 10^{6} \bigg( \frac{Q_{p0}}{m_{e}c^{2}}\bigg)^{3}\sim 10^{10} \text{ g $\cdot$ cm$^{-3}$}}.
\end{equation}
Here we consider formation of the non-equilibrium composition during cooling for the densities $\rho=10^{11}$ and $\rho=10^{13}$ g $\cdot$ cm $^{-3}$.

\center{3. NUCLEI PROPERTIES} \\ \justifying Using the finite range droplet model \cite{FRDM} as a mass model, we obtain the localization of the neutron drip line (Fig. \Ref{A(Z)}) and the proton binding energies at this boundary (Fig. \Ref{Qp(Z)_correction}). Within the framework of this model, the nucleus binding energy is described as:
\begin{equation*}
\begin{split}
&\mathrm{\Delta E(A,Z)=Z(m_{p}-m_{\mu})c^{2}+a_{3} A^{1 / 3} B_{\mathrm{k}}}\\
    &\mathrm{+(A-Z)(m_{n}-m_{\mu})c^{2}+a_{0}A^{0}+c_{1} \frac{Z^{2}}{A^{1 / 3}} B_{3}}\\
    &\mathrm{+\left(-a_{1}+J \bar{\delta}^{2}-\frac{1}{2} K \bar{\varepsilon}^{2}\right)A- c_{2} Z^{2} A^{1 / 3} B_{\mathrm{r}}}\\
    &\mathrm{+\left(a_{2} B_{1}+\frac{9}{4} \frac{J^{2}}{Q} \bar{\delta}^{2} \frac{B_{\mathrm{s}}^{2}}{B_{1}}\right) A^{2 / 3}-c_{4} \frac{Z^{4 / 3}}{A^{1 / 3}}}
\end{split}
\end{equation*}
\begin{equation}
\label{eq:2.4.1}
\begin{split}
&\mathrm{+W \left( |I|+\left\{\begin{array}{ll}
1 / \mathrm{A}, & \mathrm{Z}=\mathrm{N} \text { and both odd} \\
0, & \text {even }
\end{array}\right)\right.}\\
    &\mathrm{-c_{5} Z^{2} \frac{B_{\mathrm{W}} B_{\mathrm{s}}}{B_{1}}-c_{\mathrm{a}}(N-Z)+f_{0} \frac{Z^{2}}{A}-a_{\mathrm{el}} Z^{2.39}}\\
    &\mathrm{+\left\{\begin{array}{ll}
+\bar{\Delta}_{\mathrm{p}}+\bar{\Delta}_{\mathrm{n}}-\delta_{\mathrm{np}}, & \mathrm{Z} \text { and } \mathrm{N} \text { odd} \\
+\bar{\Delta}_{\mathrm{p}},&\mathrm{Z} \text { odd and } \mathrm{N} \text { even} \\
+\bar{\Delta}_{\mathrm{n}},&\mathrm{Z} \text { even and } \mathrm{N} \text { odd} \\
+0, & \mathrm{Z} \text { and } \mathrm{N} \text { even},
\end{array}\right.}
\end{split}
\end{equation}
where $\mathrm{m_{\mu}=1.660539 \cdot 10^{-24}}$ g is the atomic mass unit. $\mathrm{B_{i}}$ -- functions that take into account the finite range of nuclear forces, the shape of the nucleus, relative Coulomb energy, curvature, surface and volume energies; $\mathrm{I=(A-2Z)/A}$ -- relative neutron excess; $\mathrm{a_{i}, \;c_{i}, \;J,  \;Q, \;L}$, C, $\gamma$, W, h, $\mathrm{r_{max}}$, $\mathrm{f_{0}}$ -- constants; parameters $\mathrm{\bar{\Delta}_{n}, \; \bar{\Delta}_{p}, \; \bar{\varepsilon}, \; \bar{\delta}}$ and $\mathrm{\delta_{np}}$ -- functions that take into account the even-odd nuclei properties. A detailed description of each function can be found in the \cite{FRDM} paper.

\center{4. NUMERICAL SIMULATION} \\ \justifying A Fortran program was developed to simulate the cooling of a hot matter. The explicit Runge-Kutta's fourth order method with automatic step selection was used. For numerical calculation, it is convenient to use dimensionless expressions of the involved values. Introducing concentrations, the change in each isotope due to nuclear and beta reactions is written as follows:
\begin{equation*}
\begin{split}
    &\mathrm{\frac{dY(A,Z)}{dt}}=-\lambda_{\beta-}(\mathrm{A}, \mathrm{Z}) \mathrm{Y}(\mathrm{A}, \mathrm{Z})\\
    &+\lambda_{\beta-}(\mathrm{A}, \mathrm{Z}-1) \mathrm{Y}(\mathrm{A}, \mathrm{Z}-1)\\
    &-\lambda_{\beta+}(\mathrm{A},\mathrm{Z})\mathrm{Y}(\mathrm{A},\mathrm{Z})\\
    &+\lambda_{\beta+}(\mathrm{A}, \mathrm{Z}+1) \mathrm{Y}(\mathrm{A}, \mathrm{Z}+1)\\
    &-\lambda_{\text {cap } \beta-}(\mathrm{A}, \mathrm{Z})\mathrm{Y}(\mathrm{A},\mathrm{Z})\\
    &+\lambda_{\text {cap } \beta-}(\mathrm{A}, \mathrm{Z}+1) \mathrm{Y}(\mathrm{A}, \mathrm{Z}+1)\\
    &-\lambda_{\text {cap }\beta+}(\mathrm{A},\mathrm{Z})\mathrm{Y}(\mathrm{A},\mathrm{Z})\\
    &+\lambda_{\mathrm{cap} \beta+}(\mathrm{A}, \mathrm{Z}-1) \mathrm{Y}(\mathrm{A}, \mathrm{Z}-1)\\
    &-\lambda_{\mathrm{n\gamma}}(\mathrm{A},\mathrm{Z})\mathrm{Y}(\mathrm{A},\mathrm{Z})\\
    &+\lambda_{\mathrm{n} \gamma}(\mathrm{A}-1, \mathrm{Z}) \mathrm{Y}(\mathrm{A}-1, \mathrm{Z})
\end{split}
\end{equation*}
\begin{equation}
\label{eq:3.1}
\begin{split}
&-\lambda_{\gamma \mathrm{n}}(\mathrm{A}, \mathrm{Z}) \mathrm{Y}(\mathrm{A},\mathrm{Z})\\
    &+\lambda_{\gamma \mathrm{n}}(\mathrm{A}+1, \mathrm{Z}) \mathrm{Y}(\mathrm{A}+1, \mathrm{Z})\\
&-\lambda_{\mathrm{p} \gamma}(\mathrm{A},\mathrm{Z})\mathrm{Y}(\mathrm{A},\mathrm{Z})\\
    &+\lambda_{\mathrm{p} \gamma}(\mathrm{A}-1, \mathrm{Z}-1) \mathrm{Y}(\mathrm{A}-1, \mathrm{Z}-1)\\
    &-\lambda_{\gamma \mathrm{p}}(\mathrm{A}, \mathrm{Z}) \mathrm{Y}(\mathrm{A},\mathrm{Z})\\
    &+\lambda_{\gamma \mathrm{p}}(\mathrm{A}+1, \mathrm{Z}+1) \mathrm{Y}(\mathrm{A}+1, \mathrm{Z}+1).
\end{split}
\end{equation}
Here we denote the reduced concentrations as $\mathrm{Y}_{\mathrm{i}}=\mathrm{n}_{\mathrm{i}}/ \left(\rho \cdot \mathrm{N}_{\mathrm{A}} \right)$, $\mathrm{N_{A}}$ is Avogadro's number. The two-particle reactions $\mathrm{i}^{\eta} (\mathrm{j}, \mathrm{\gamma}) \mathrm{m}^{\nu}$ rates $\lambda_{\mathrm{j \gamma}}$, where $\mathrm{j=n,p}$, and $\eta$ and $\nu$ take into account the states of the initial and final nuclei $\mathrm{\textsl{i}}$ and $\mathrm{\textsl{m}}$, respectively, are written in the form of $\lambda_{\mathrm{j \gamma}} (\mathrm{A}, \mathrm{Z})=\mathrm{Y}_{\mathrm{j}} \cdot \rho \cdot \mathrm{N}_{\mathrm{A}} \langle \sigma_{\mathrm{i}} \mathrm{v} \rangle$, where $\mathrm{\sigma}$ is the reaction cross section and $\mathrm{v}$ is the relative velocity of the interacting particles.

All isotopes in the range $\mathrm{Z \in [0; 83]}$ with nuclear data available are initially included in the system and distributed in accordance with (\ref{Saha}) for the given initial $\mathrm{\rho}$, $\mathrm{T_{9}}$ and the ratio of the total number of protons to neutrons:
\begin{equation}
\label{eq:3.2}
\mathrm{R=\frac{n_{p}+\sum Z\cdot n(A,Z)}{n_{n}+\sum(A-Z)\cdot n(A,Z)}}.
\end{equation}

At each time step after a temperature change $\mathrm{T_{9}(t)}$ and constant density $\mathrm{\rho}$, a new chemical potential of electrons $\mathrm{\mu_{e^{-}}}$, which is equal in modulus and opposite in sign to chemical potential of positrons, is calculated. New concentrations of electrons and positrons are calculated after. In order to find them it is necessary to solve numerically the system of the equations, using Newton's method:
\begin{equation}
\label{eq:3.3}
\begin{cases}
  \mathrm{n_{e^{\pm}}=\frac{1}{\pi^{2}}\left(\frac{\mathrm{kT}}{\mathrm{c} \hbar}\right)^{3} \int\limits_{0}^{\infty} \frac{\mathrm{x}^{2} \mathrm{dx}}{1+\exp \left(\sqrt{\mathrm{x}^{2}+\alpha^{2}} - \beta_{\pm}\right)}},\\
  \mathrm{n_{e^{-}}=n_{e^{+}}+\sum\limits_{A,Z}n(A,Z)\cdot Z},
\end{cases}
\end{equation}
with respect to implicitly defined variables $\mathrm{\beta_{\pm}=\mu_{e^{\pm}} /kT}$, $n_{\pm}$, and given parameters $\mathrm{\alpha=m_{e} c^{2} /kT}$ and $\mathrm{x=cp/kT}$. At low temperatures, with strongly degenerate electron gas and the absence of positrons, the following relations \cite{ll_q_mech} have been used:
\begin{equation}
\label{eq:3.4}
\begin{cases}
  \mathrm{n_{e^{-}}=\sum\limits_{A,Z}n(A,Z)\cdot Z},\\
  \mathrm{\mu_{\mathrm{e}^{-}} \approx \varepsilon_{\mathrm{fe}}\left[1-\frac{1}{3}\left(\frac{\pi \mathrm{kT}}{\varepsilon_{\mathrm{fe}}}\right)^{2}\right]}.
\end{cases}
\end{equation}
If the chemical potentials of electrons and positrons are known, one can find the rates of beta reactions $\mathrm{W}$ and neutrino energy loss $\mathrm{\Theta}$ per nucleus \cite{BKOGAN_1989} in reactions (\ref{beta-reactions}):
\begin{equation}
\label{eq:3.5}
\begin{split}
&\mathrm{W}^{(\mathrm{a})}=\frac{\mathrm{g}_{\mathrm{z'}}}{\mathrm{g_{z}}} \frac{\ln 2}{\mathrm{Ft}_{1 / 2}}\left(\frac{\mathrm{kT}}{\mathrm{m}_{\mathrm{e}} \mathrm{c}^{2}}\right)^{5} \mathrm{I}_{2},\\
&\mathrm{W}^{(\mathrm{b})}=\frac{\ln 2}{\mathrm{Ft}_{1 / 2}}\left(\frac{\mathrm{kT}}{\mathrm{m}_{\mathrm{e}} \mathrm{c}^{2}}\right)^{5} \mathrm{I}_{2}^{\prime},\\
&\mathrm{\mathrm{\Theta}^{(\mathrm{a})}=\frac{\mathrm{g}_{\mathrm{z'}}}{\mathrm{g}_{\mathrm{z}}} \frac{\ln 2}{\mathrm{Ft}_{1 / 2}}\left(\frac{\mathrm{kT}}{\mathrm{m}_{\mathrm{e}} \mathrm{c}^{2}}\right)^{6} \mathrm{m}_{\mathrm{e}} \mathrm{c}^{2} \mathrm{I}_{3}},\\
&\mathrm{\Theta}^{(\mathrm{b})}=\frac{\ln 2}{\mathrm{Ft}_{1 / 2}}\left(\frac{\mathrm{kT}}{\mathrm{m}_{\mathrm{e}} \mathrm{c}^{2}}\right)^{6} \mathrm{m}_{\mathrm{e}} \mathrm{c}^{2} \mathrm{I}_{3}^{\prime}.
\end{split}
\end{equation}
The integrals $\mathrm{I_{k}}$ and $\mathrm{I'_{k}}$ are defined as:
\begin{equation}
\label{eq:3.6}
\begin{split}
& \mathrm{I_{k}=\int\limits_{0}^{\infty} \frac{x^{k}\left(x+x_{0}\right) \sqrt{\left(x+x_{0}\right)^{2}-\alpha^{2}}}{1+\exp \left(x+x_{0}-\beta_{\pm}\right)} d x}, \\
& \mathrm{I_{k}^{\prime}=\int\limits_{0}^{x_{0}-\alpha} \frac{\left(x_{0}-\alpha-x\right)^{k}(x+\alpha) \sqrt{x^{2}+2 \alpha x}}{1+\exp \left(\beta_{\pm}-x-\alpha\right)} d x},
\end{split}
\end{equation}
where we denote $\mathrm{x_{0}=c^{2}[m(A,Z)-m(A,Z')]/kT}$; $\mathrm{g_i}$ are statistical weights. The constants $\mathrm{Ft_{1/2}}$ are determined by the experimentally measured half-lives. The values of $\mathrm{Ft_{1/2}}$ differ considerably for various transitions to excited states of the same nuclei. All integral expressions in this work are calculated using the Gauss method and the similar ones, utilizing from 5 up to 11 points with the corresponding weights \cite{gausse,num_recipe77}. In the case of strong degeneracy at low temperatures, the integral expressions (\ref{eq:3.5}) turn to the following ones \cite{BKOGAN_1989}:
\begin{equation}
\label{15}
\begin{split}
&\mathrm{W}^{(\mathrm{a})}=\frac{\mathrm{g}_{\mathrm{z'}}}{\mathrm{g_{z}}} \frac{\ln 2}{\mathrm{Ft}_{1 / 2}} \mathrm{\int\limits_{\delta}^{u_{fe}} u\sqrt{u^{2}-1}(u-\delta)^{2}du},\\
&\mathrm{W}^{(\mathrm{b})}=\frac{\ln 2}{\mathrm{Ft}_{1 / 2}} \mathrm{\int\limits_{u_{fe}}^{\delta} u\sqrt{u^{2}-1}(\delta-u)^{2}du},\\
&\mathrm{\mathrm{\Theta}^{(\mathrm{a})}=\frac{\mathrm{g}_{\mathrm{z'}}}{\mathrm{g}_{\mathrm{z}}} \frac{\ln 2}{\mathrm{Ft}_{1 / 2}} \mathrm{m}_{\mathrm{e}} \mathrm{c}^{2}\mathrm{\int\limits_{\delta}^{u_{fe}} u\sqrt{u^{2}-1}(u-\delta)^{3}du}},\\
&\mathrm{\mathrm{\Theta}^{(\mathrm{b})}=\frac{\ln 2}{\mathrm{Ft}_{1 / 2}} \mathrm{m}_{\mathrm{e}} \mathrm{c}^{2}\mathrm{\int\limits_{u_{fe}}^{\delta} u\sqrt{u^{2}-1}(\delta-u)^{3}du}},
\end{split}
\end{equation}
which are calculated analytically \cite{BKOGAN_1989,fk1962}.

An approximate equation for the non-equilibrium layer cooling is used in the following form:
\begin{equation}
\label{eq:3.7}
\mathrm{\frac{3}{2}k \frac{dT}{dt}\cdot \sum\limits_{(A,Z)}n(A,Z)= \sum\limits_{(A,Z)}\Theta(A,Z)\cdot n(A,Z)}.
\end{equation}
Only the volume energy losses of the non-equilibrium layer due to neutrino radiation are taken into account. Considering the heating of matter due to non-equilibrium beta processes, one can obtain a self-consistent case in which the rates of neutrino energy losses are compensated by the non-equilibrium heating. Accurate accounting for non-equilibrium heating of matter is not considered in this work. It is assumed that non-equilibrium heating compensates for neutrino energy losses at a temperature of $\mathrm{T_{9}=0.1}$ and the matter is not cooled rapidly further.

At each time step with a constant density $\mathrm{\rho}$, beta reactions change the temperature $\mathrm{T_{9}(t)}$ and the ratio of the total number of protons to neutrons $\mathrm{R(t)}$. With a new $\mathrm{T}$ and $\mathrm{R}$, considering nuclear statistical equilibrium conditions (\ref{Saha}), the implicit equations (\ref{Saha}) and (\ref{eq:3.3}) are solved simultaneously, using Newton's method. New nuclear composition, values $\mathrm{\beta_{\pm}=\mu_{e^{-}}/kT}$ and $n_{\pm}$ are found.
When the temperature drops to $\mathrm{T_{9}\sim 3-5}$ and below, the equilibrium (\ref{Saha}) is violated due to the Coulomb barrier. Equilibrium for protons and neutrons (\ref{Saha}) is replaced with equilibrium for neutrons only. Similarly to the Saha expression (\ref{Saha}), one derive equation for adjacent atomic mass numbers $\mathrm{A}$ in isotopic chain with fixed atomic charge numbers $\mathrm{Z}$ as function of free neutrons concentration:
\begin{equation}
\label{SahaN}
\begin{split}
&\mathrm{\frac{n(A,Z)}{n(A+k,Z)}=\bigg(\frac{A}{A+k}\bigg)^{\frac{3}{2}}\bigg(\frac{2}{n_{n}}\bigg)^{k}}\\
&\mathrm{\times \bigg( \frac{m_{n} kT}{2\pi \hbar^{2}} \bigg)^{\frac{3}{2}k} \frac{g(A,Z)}{g(A+k,Z)}} \\
&\mathrm{\times \exp \bigg( \frac{Q(A,Z)-Q(A+k,Z)}{kT} \bigg) }.
\end{split}
\end{equation}
The criterion for terminating the calculation is that the concentrations of elements and most parameters reach constant values. It occurs in $\mathrm{t \sim 2 \cdot 10^{2}} $ s for the problem considered.

Degenerate electrons make the key contribution to the pressure for the chosen conditions. We evaluate the pressure according to \cite{zeldovich} in the following form:
\begin{equation}
\label{eq:2.6.1}
\mathrm{P\approx 1.2 \cdot 10^{15} \bigg( \frac{\rho}{\mu_{z}}\bigg)^{4/3}}.
\end{equation}
A pressure difference at the boundaries of the non-equilibrium layer between the densities $\mathrm{\rho_{max}}$ and $\mathrm{\rho_{min}}$ is:
\begin{equation}
\label{eq:2.6.2}
\mathrm{\Delta P=P_{max}-P_{min}\sim 10^{31} \text{ Ba}}.
\end{equation}
A mass evaluation of the non-equilibrium layer in the crust of the neutron star envelope $\mathrm{M_{shell}}$ is found using the forces balance equation in the envelope of a neutron star with mass $\mathrm{M_{NS}}$. That gives in a flat layer approximation the following estimation: \cite{B-kogan1}
\begin{equation}
\label{eq:2.6.3}
\mathrm{M_{shell}=\frac{4 \pi R^{4}}{GM_{NS}}\Delta P\approx 0.1 \Delta P \sim 10^{30} \text{ g}}.
\end{equation}

To evaluate the nuclear energy stored in the non-equilibrium layer relative to the equilibrium nuclear composition let us assume that $\mathrm{\Delta E \sim 7 \cdot 10^{-3} \, m_{n} c^2}$ energy is released per nucleon \cite{B-kogan1} during the transition to equilibrium.
For the total energy stored we get:
\begin{equation}
\label{eq:2.6.6}
\mathrm{E\sim\frac{M_{shell}}{m_\mu} \cdot \Delta E \sim 10^{48} \text{ erg}}.
\end{equation}
The presence of such large nuclear energy reserves in the non-equilibrium layer is sufficient to maintain the X-ray luminosity at the level of $\mathrm{L=10^{36-37}}$ erg/s for tens of thousands of years, and can also lead to powerful nuclear explosions accompanying starquakes of neutron stars \cite{bkinch,bkch81,bkik14}. Taking into account pycnonuclear reactions that increase the upper density boundary of the non-equilibrium layer by about an order of magnitude \cite{sato}, as well as lower mass neutron stars with a thicker non-equilibrium layer \cite{bkik14}, the nuclear energy reserve in the non-equilibrium layer increases by more than an order of magnitude.
\begin{figure*}[ht!]
  \includegraphics[width=\textwidth,height=0.38\textheight]{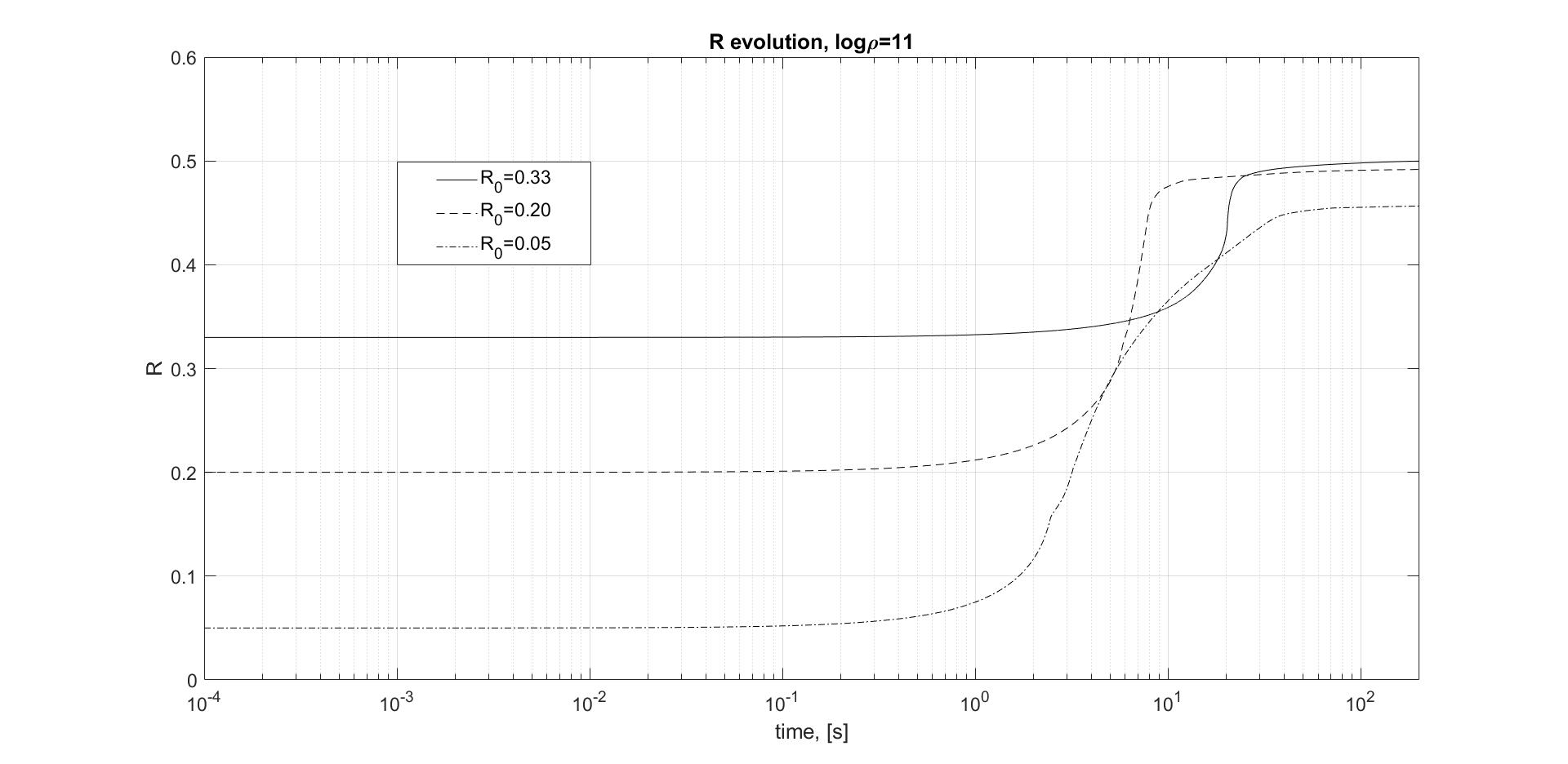}
  \caption{Time evolution of different parameter $\mathrm{R}$ values at density of $\mathrm{log\rho=11}$ during cooling.}
  \label{R-10-11}
\end{figure*}

As a result of various neutron star activity processes, such as starquakes, the matter of the non-equilibrium layer can be moved from inner layers into the outer ones. Under conditions of low density, the degree of electron degeneracy decreases, and the Pauli principle does not restrict beta reactions anymore, during the time scale of which the stored nuclear energy can be released in an explosive manner. The trigger of such an explosion can play nuclear reactions, considered in \cite{bkch83}, in the ejected part of the superheavy nuclei matter that lead to chain nuclear fission, followed by the explosion.

\begin{figure*}[ht!]
  \includegraphics[width=\textwidth,height=0.38\textheight]{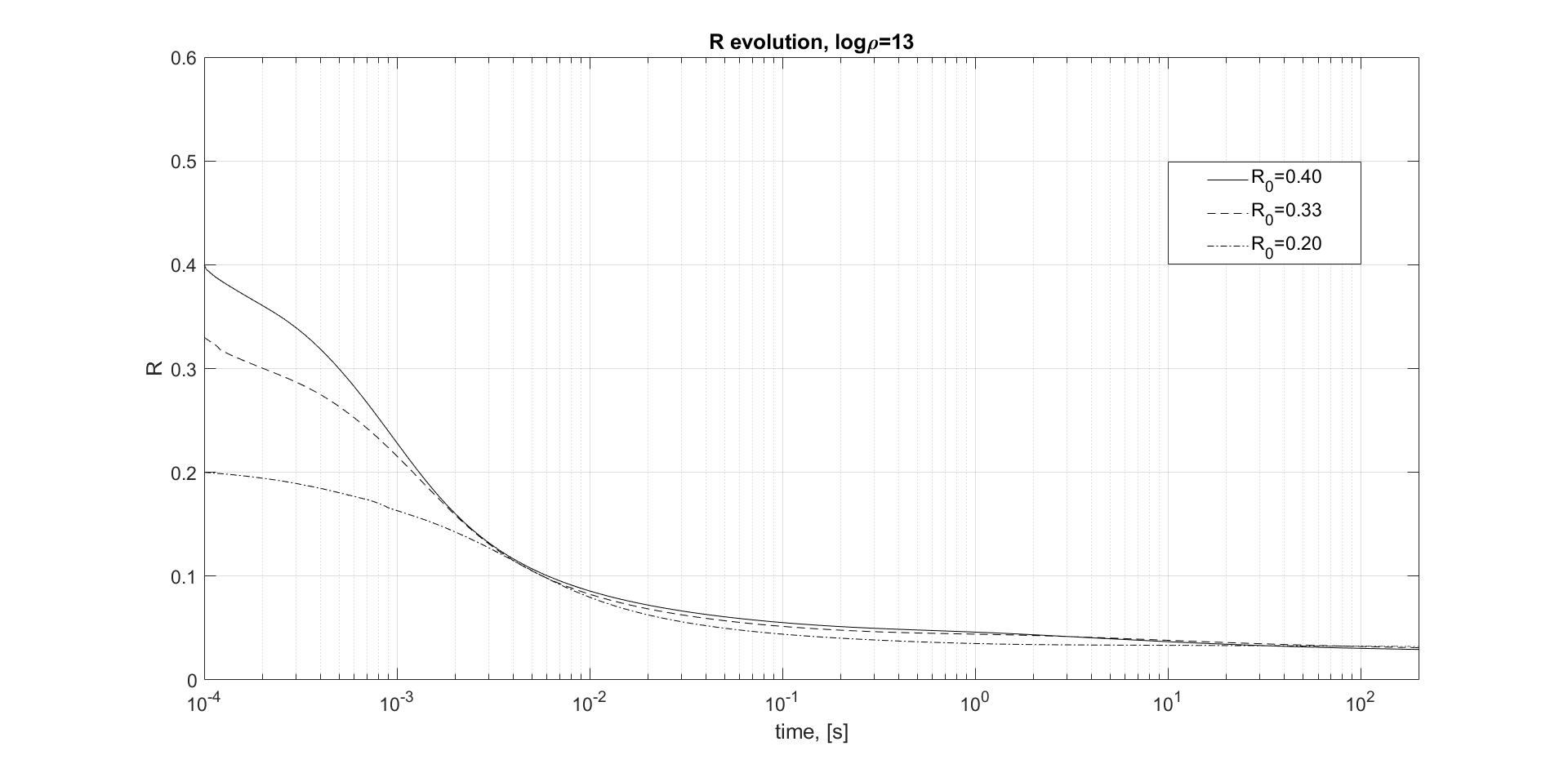}
  \caption{Time evolution of different parameter $\mathrm{R}$ values at density of $\mathrm{log\rho=13}$ during cooling.}
  \label{R-10-13}
\end{figure*}

\center{5. RESULTS AND DISCUSSION} \\ \justifying Let us consider the results of a numerical calculation of the chemical composition evolution. The initial temperature for all calculating tracks is $\mathrm{T_9=10}$. For each density value $\mathrm{log \rho = 11,\,13}$, several values of the $\mathrm{R}$ parameter are selected. Only the most common elements with mass fractions $\mathrm{x_{j}>10^{-4}}$ are presented on the graphs that display the initial and final isotope distributions. Figures \ref{R-10-11} and \ref{R-10-13} correspond to different trajectories of the parameter $\mathrm{R}$ at densities $\mathrm{log \rho=11} $ and $\mathrm{log \rho=13}$ respectively.

\begin{figure*}[hb!]
  \includegraphics[width=\textwidth,height=0.38\textheight]{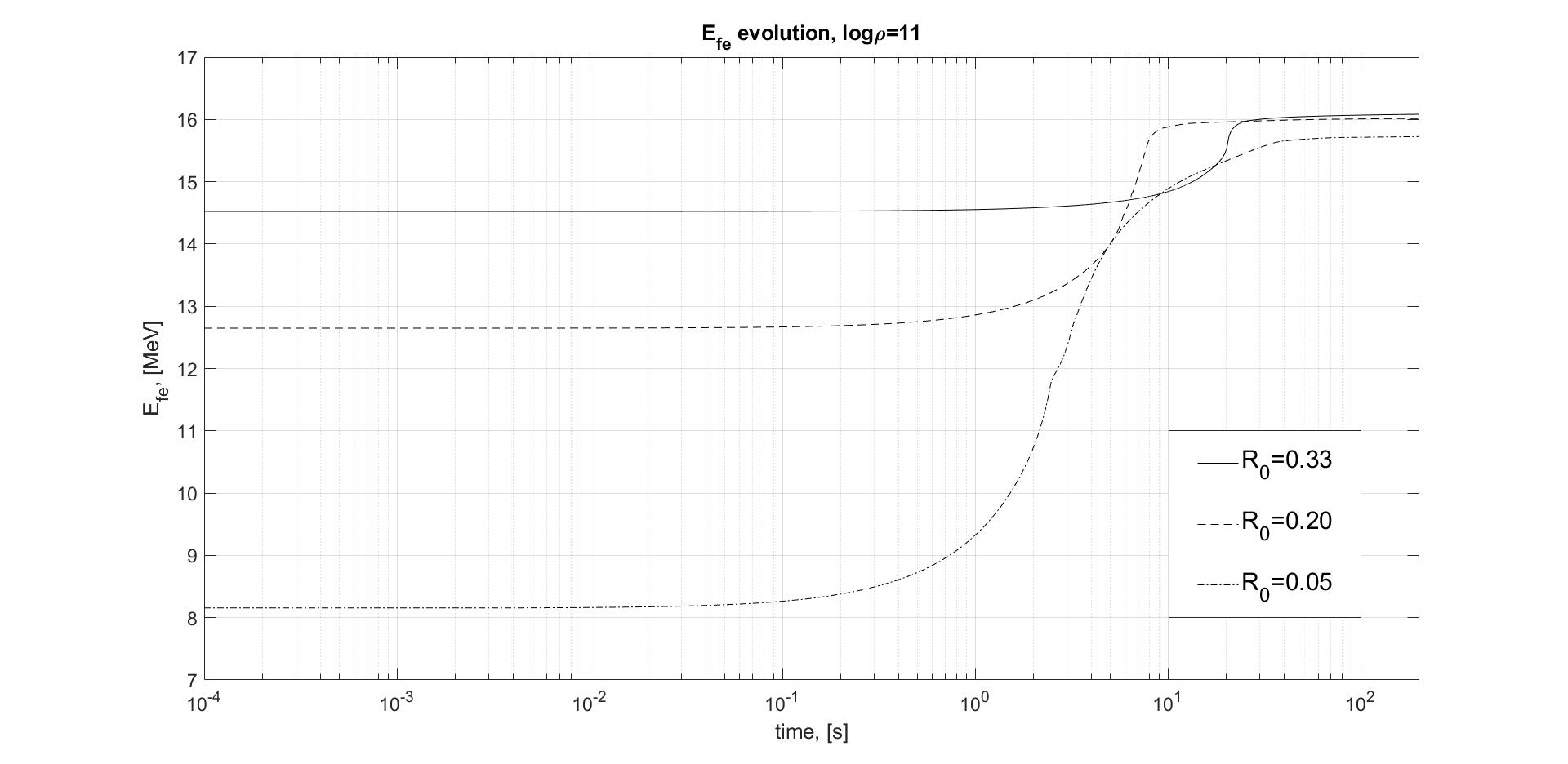}
  \caption{Time evolution of the Fermi energies $\mathrm{\varepsilon_{fe}}$ corresponding to different parameter $\mathrm{R}$ initial values at density of $\mathrm{log \rho=11}$ during cooling.}
  \label{E_fem-10-11}
\end{figure*}

\begin{figure*}[ht!]
  \includegraphics[width=\textwidth,height=0.38\textheight]{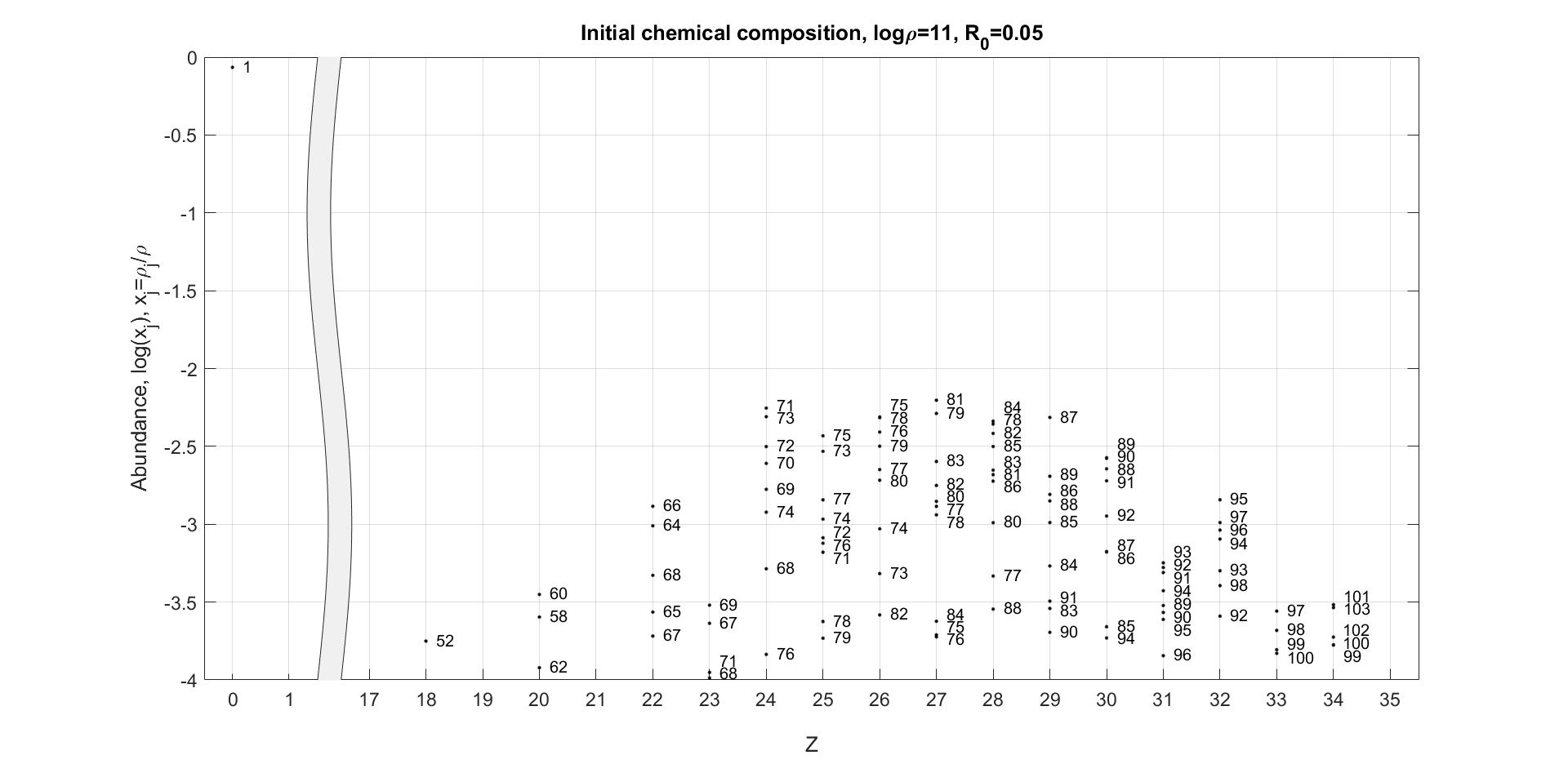}
   \caption{Initial chemical composition at $\mathrm{log \rho=11}$, $\mathrm{T_{9}=10}$ and $\mathrm{R_{0}=0.05}$. Each isotope is marked by the atomic mass number $\mathrm{A}$.}
  \label{Initial-10-11-05}
\end{figure*}

\begin{figure*}[hb!]
    \includegraphics[width=\textwidth,height=0.38\textheight]{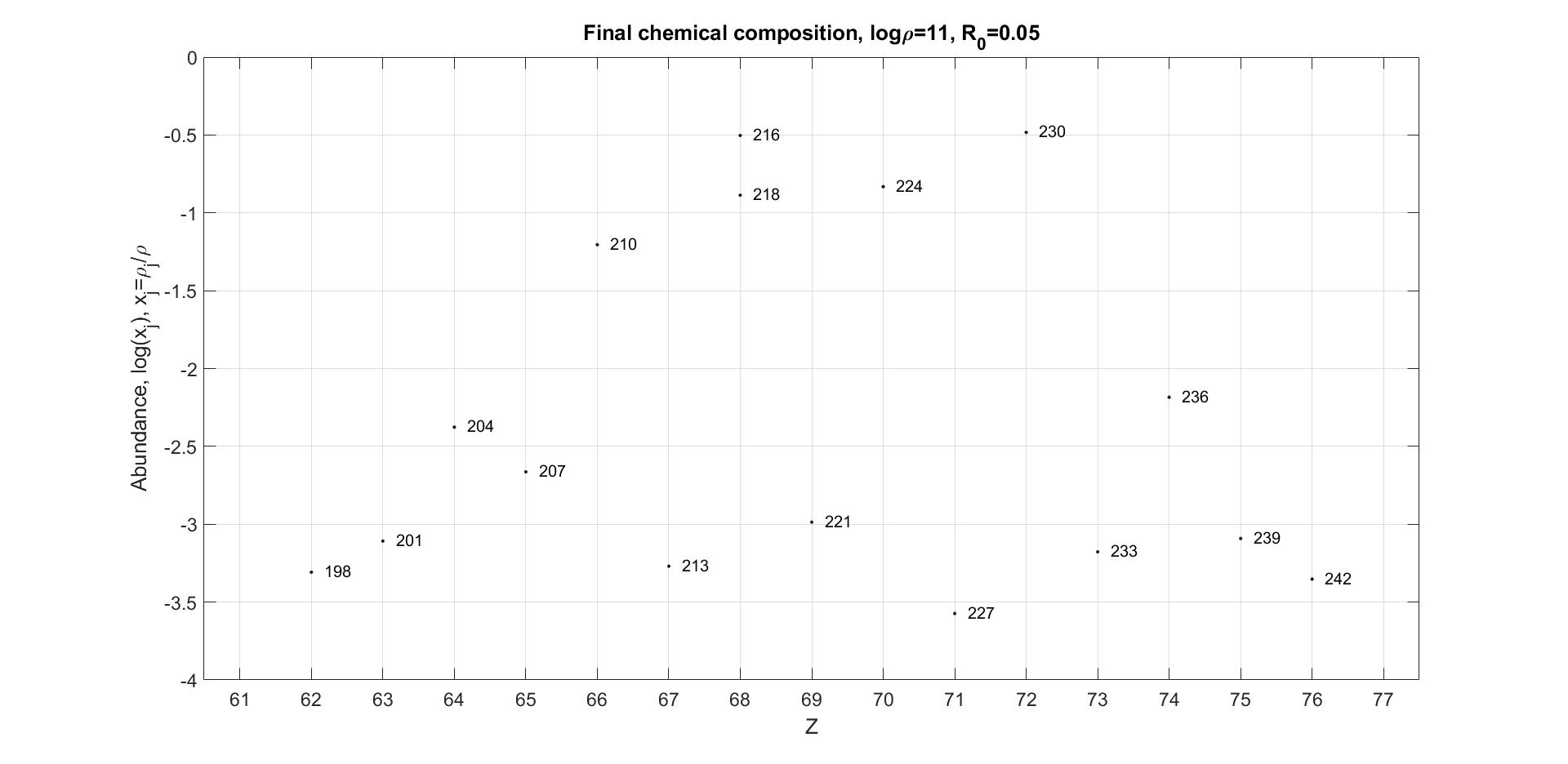}
\caption{Chemical composition at the last calculated point $\mathrm {t = 200}$ s for $\mathrm{log \rho=11}$, $\mathrm{T_{9}=10}$ and $\mathrm{R_{0}=0.05}$. Each isotope is marked by the atomic mass number $\mathrm{A}$.}
  \label{Final-10-11-05}
\end{figure*}

\begin{figure*}[ht!]
\includegraphics[width=\textwidth,height=0.38\textheight]{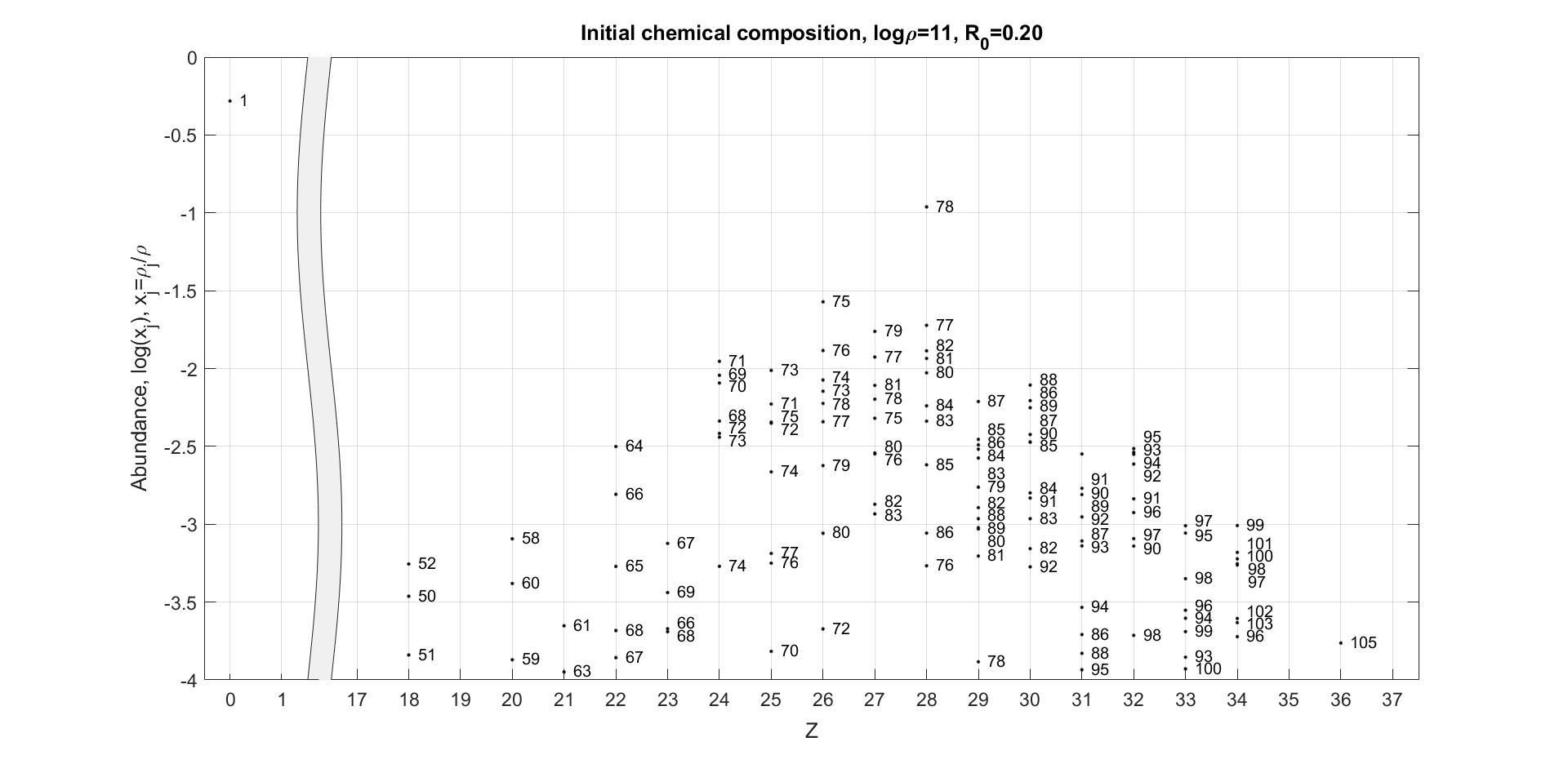}
   \caption{Initial chemical composition at $\mathrm{log \rho=11}$, $\mathrm{T_{9}=10}$ and $\mathrm{R_{0}=0.20}$. Each isotope is marked by the atomic mass number $\mathrm{A}$.}
  \label{Initial-10-11-20}
\end{figure*}

\begin{figure*}[hb!]
    \includegraphics[width=\textwidth,height=0.38\textheight]{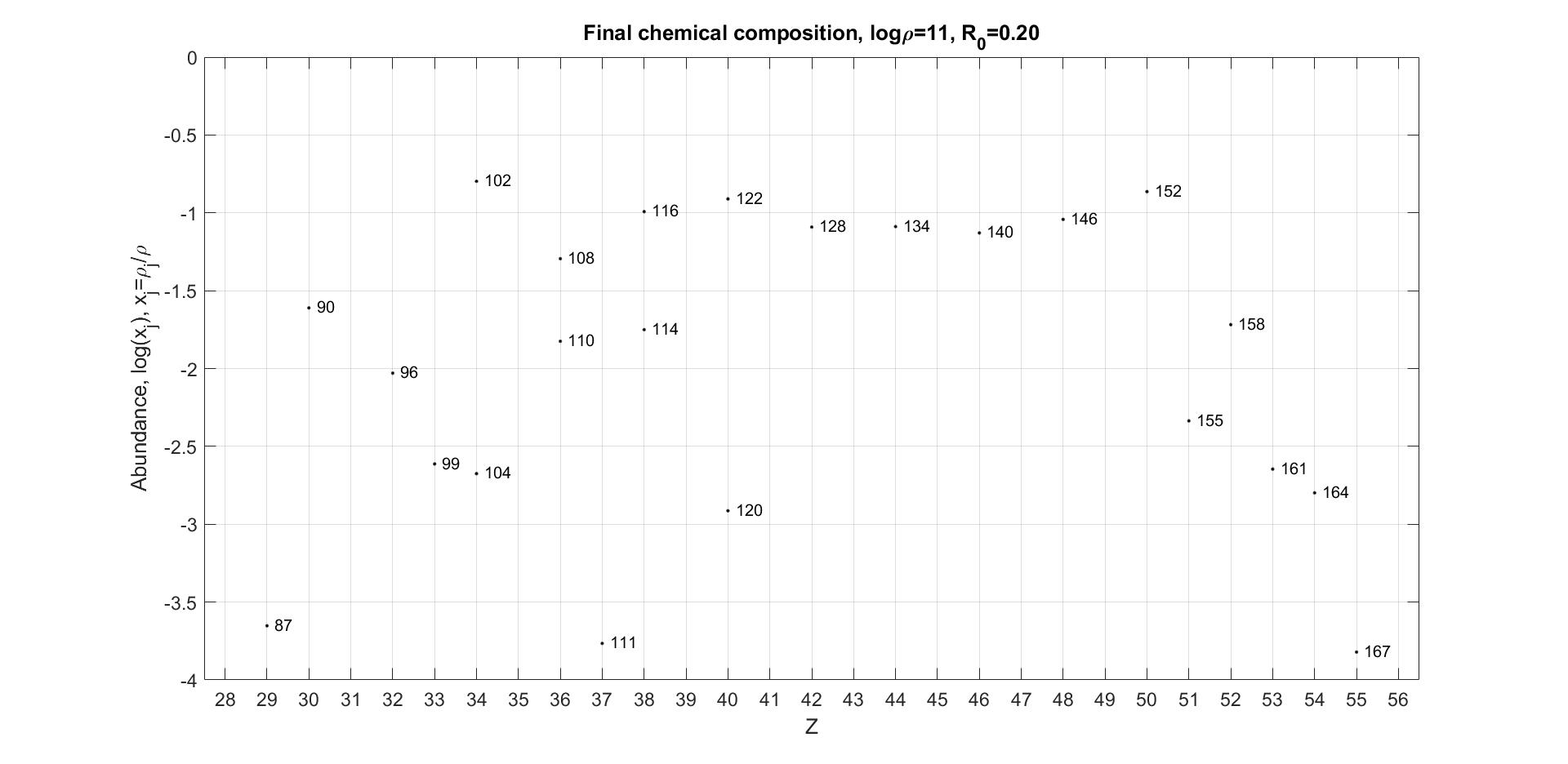}
   \caption{Chemical composition at the last calculated point $\mathrm {t = 200}$ s for $\mathrm{log \rho=11}$, $\mathrm{T_{9}=10}$ and $\mathrm{R_{0}=0.20}$. Each isotope is marked by the atomic mass number $\mathrm{A}$.}
  \label{Final-10-11-20}
\end{figure*}

\begin{figure*}[ht!]
\includegraphics[width=\textwidth,height=0.38\textheight]{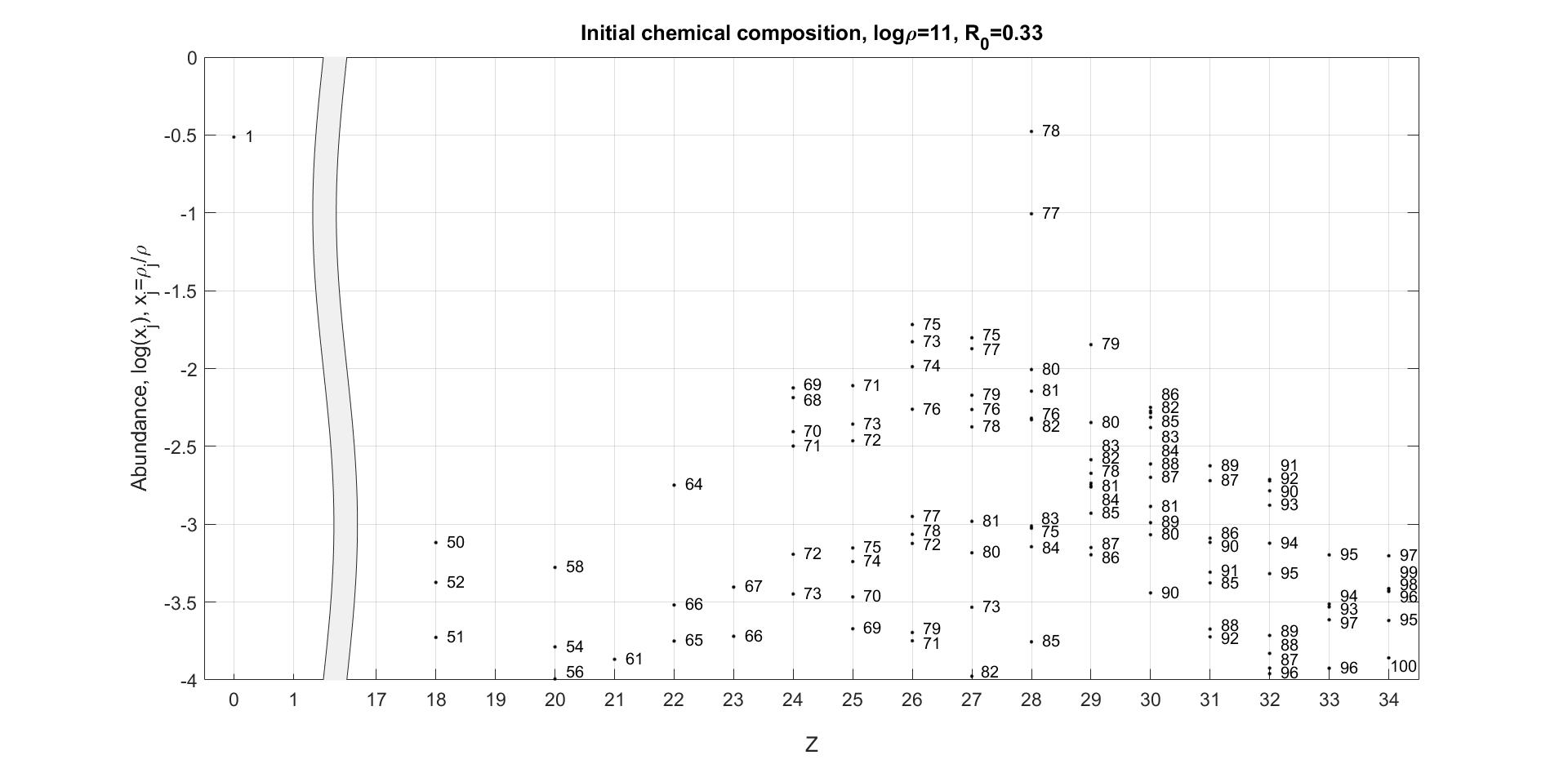}
   \caption{Initial chemical composition at $\mathrm{log \rho=11}$, $\mathrm{T_{9}=10}$ and $\mathrm{R_{0}=0.33}$. Each isotope is marked by the atomic mass number $\mathrm{A}$.}
  \label{Initial-10-11-33}
\end{figure*}

\begin{figure*}[hb!]
    \includegraphics[width=\textwidth,height=0.38\textheight]{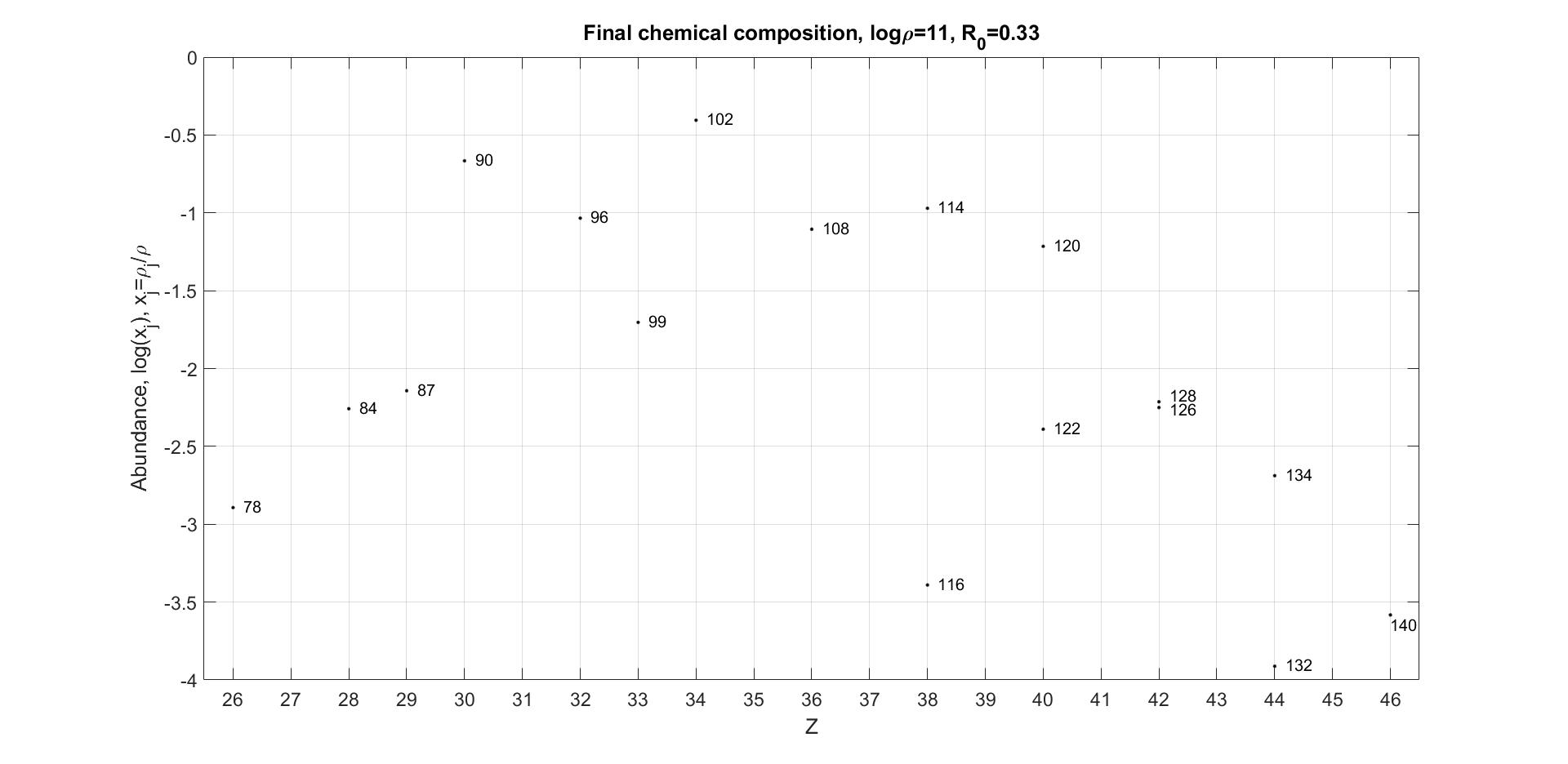}
   \caption{Chemical composition at the last calculated point $\mathrm {t = 200}$ s for $\mathrm{log \rho=11}$, $\mathrm{T_{9}=10}$ and $\mathrm{R_{0}=0.33}$. Each isotope is marked by the atomic mass number $\mathrm{A}$.}
  \label{Final-10-11-33}
\end{figure*}

\begin{figure*}[hb!]
  \includegraphics[width=\textwidth,height=0.38\textheight]{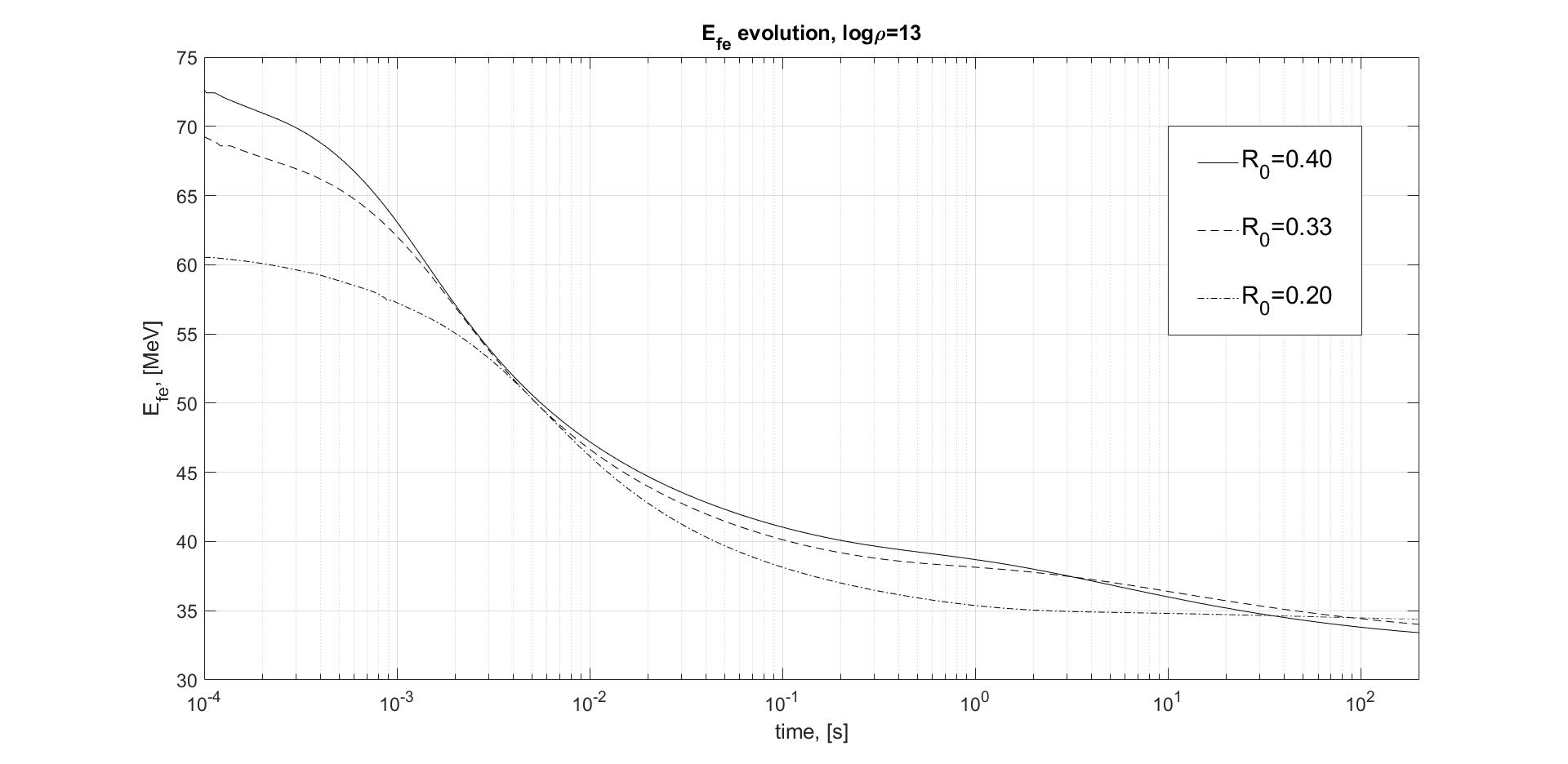}
  \caption{Time evolution of the Fermi energies $\mathrm{\varepsilon_{fe}}$ corresponding to different parameter $\mathrm{R}$ initial values at density of $\mathrm{log \rho=13}$ during cooling.}
  \label{E_fem-10-13}
\end{figure*}

\begin{figure*}[ht!]
  \includegraphics[width=\textwidth,height=0.38\textheight]{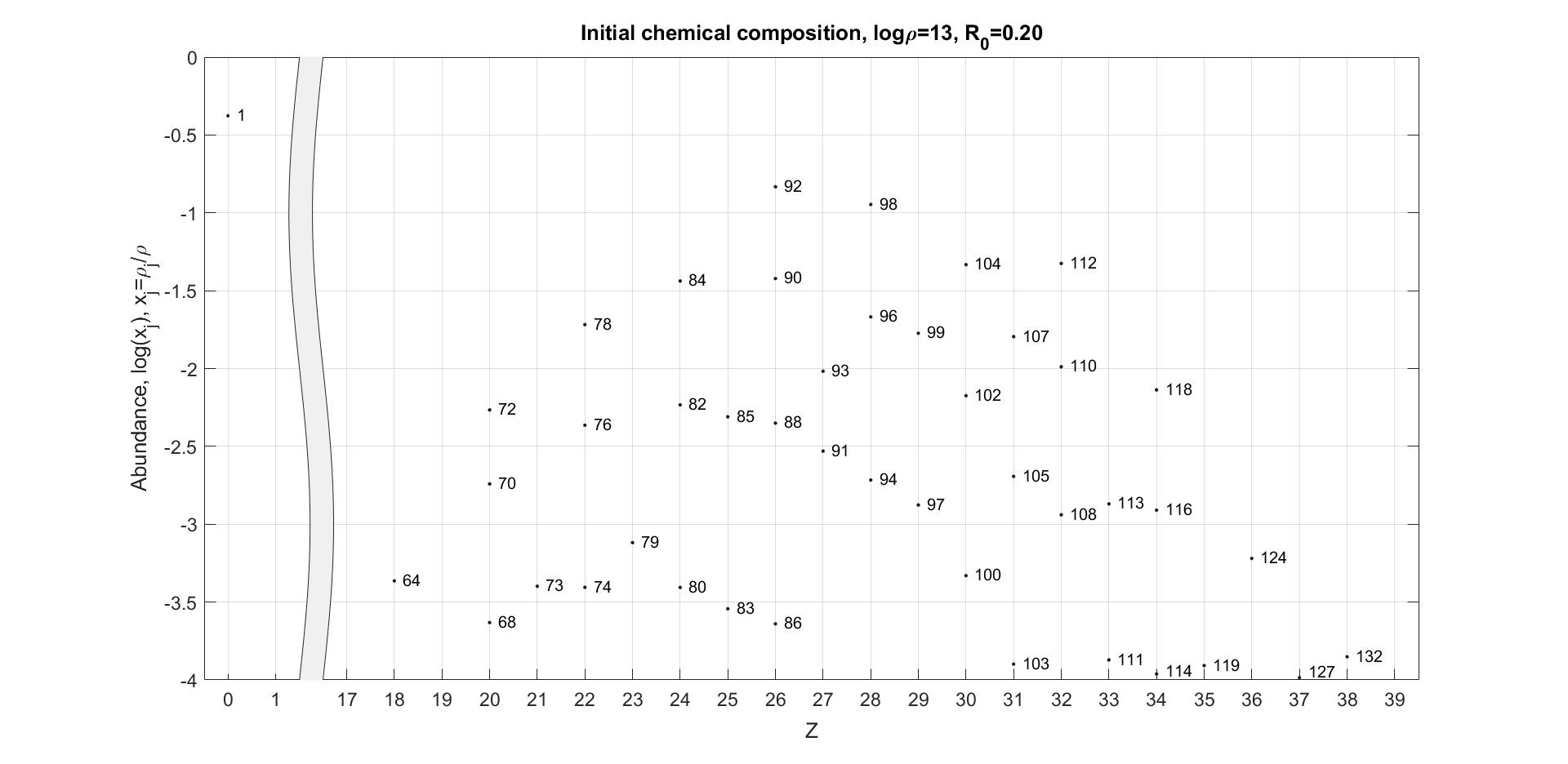}
   \caption{Initial chemical composition at $\mathrm{log \rho=13}$, $\mathrm{T_{9}=10}$ and $\mathrm{R_{0}=0.20}$. Each isotope is marked by the atomic mass number $\mathrm{A}$.}
  \label{Initial-10-13-20}
\end{figure*}

\begin{figure*}[hb!]
    \includegraphics[width=\textwidth,height=0.38\textheight]{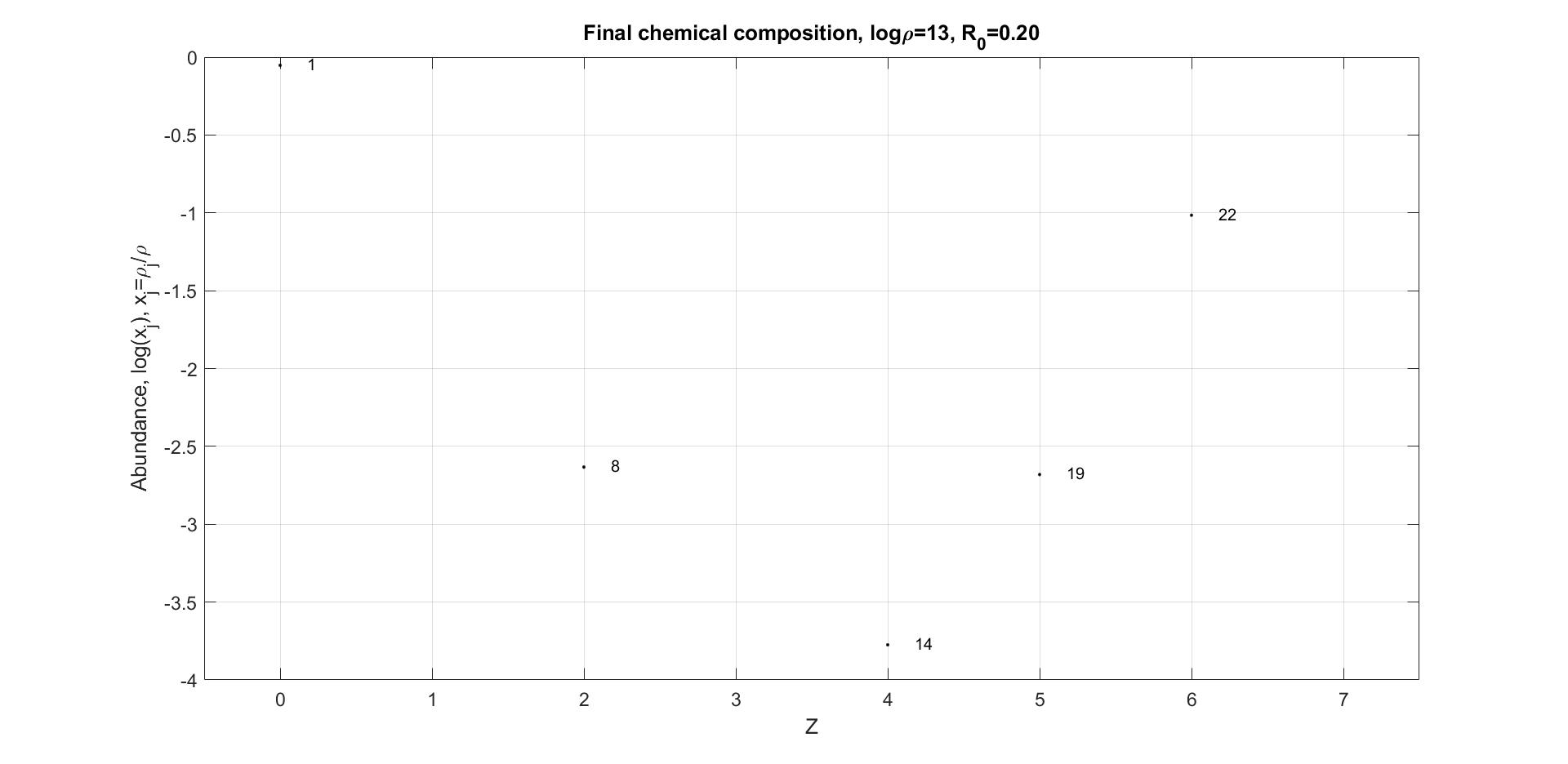}
   \caption{Chemical composition at the last calculated point $\mathrm {t = 200}$ s for $\mathrm{log \rho=13}$, $\mathrm{T_{9}=10}$ and $\mathrm{R_{0}=0.20}$. Each isotope is marked by the atomic mass number $\mathrm{A}$.}
  \label{Final-10-13-20}
\end{figure*}

\begin{figure*}[ht!]
\includegraphics[width=\textwidth,height=0.38\textheight]{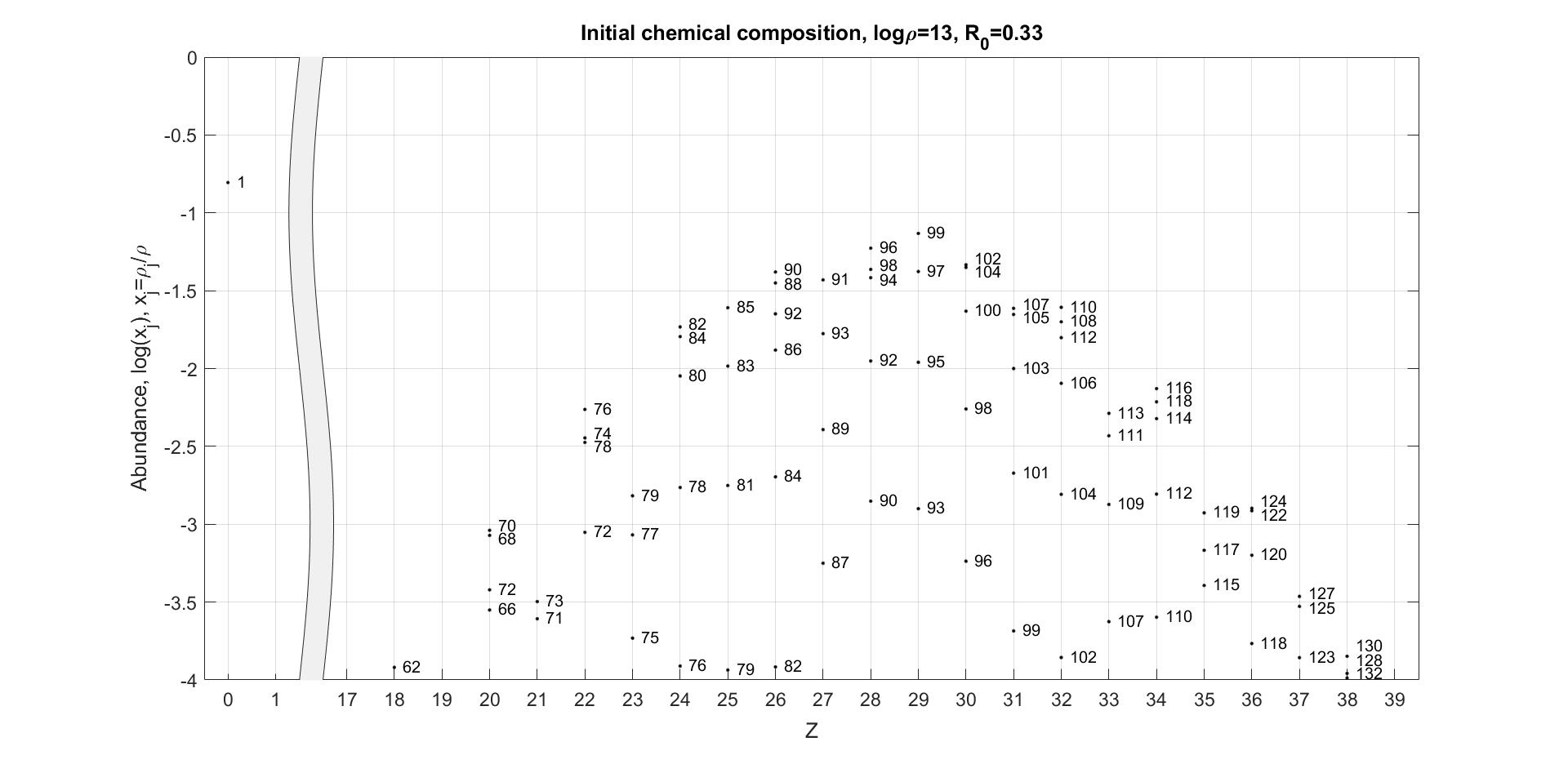}
   \caption{Initial chemical composition at $\mathrm{log \rho=13}$, $\mathrm{T_{9}=10}$ and $\mathrm{R_{0}=0.33}$. Each isotope is marked by the atomic mass number $\mathrm{A}$.}
  \label{Initial-10-13-33}
\end{figure*}

\begin{figure*}[hb!]
    \includegraphics[width=\textwidth,height=0.38\textheight]{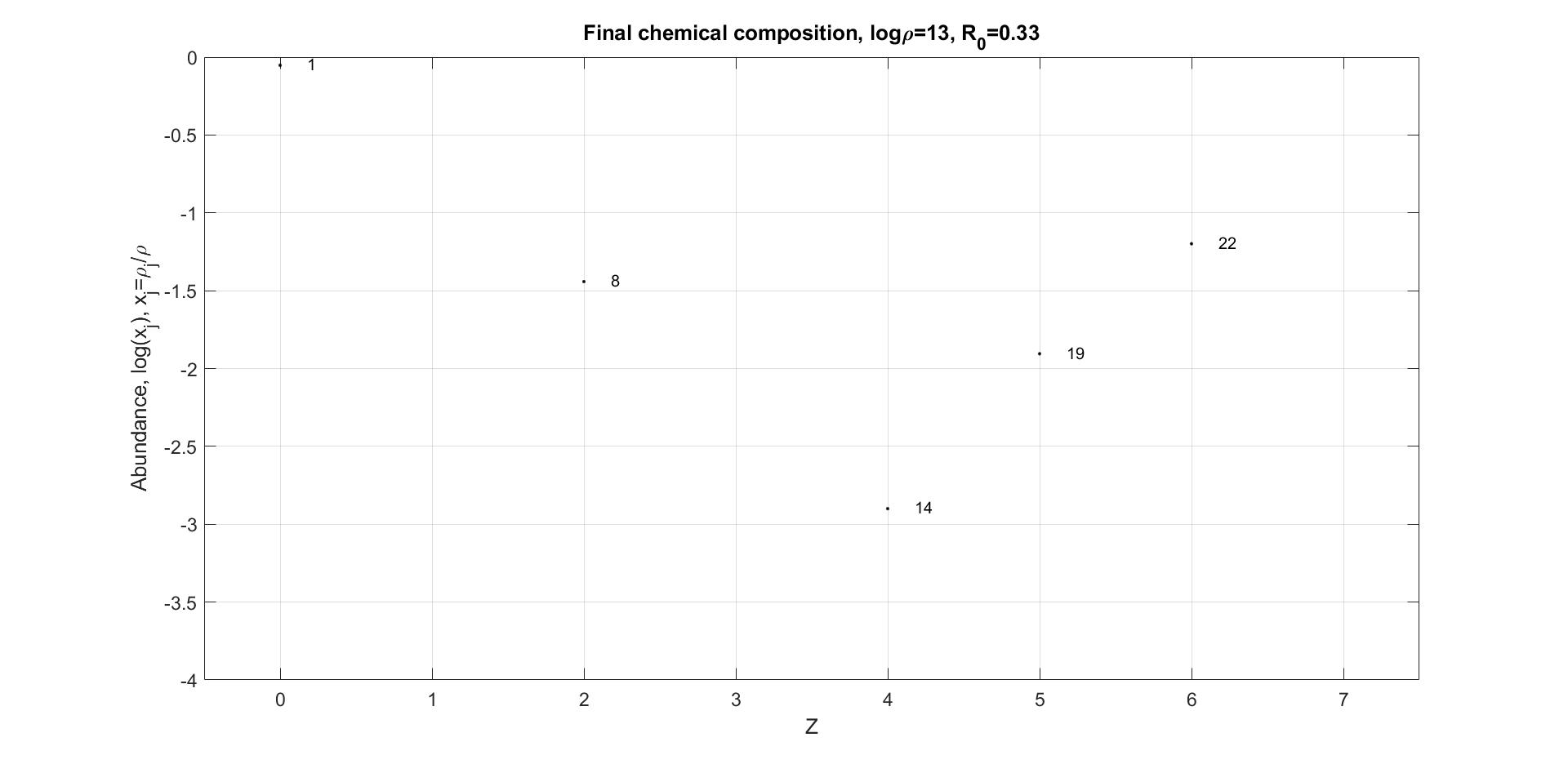}
   \caption{Chemical composition at the last calculated point $\mathrm {t = 200}$ s for $\mathrm{log \rho=13}$, $\mathrm{T_{9}=10}$ and $\mathrm{R_{0}=0.33}$. Each isotope is marked by the atomic mass number $\mathrm{A}$.}
  \label{Final-10-13-33}
\end{figure*}

\begin{figure*}[ht!]
\includegraphics[width=\textwidth,height=0.38\textheight]{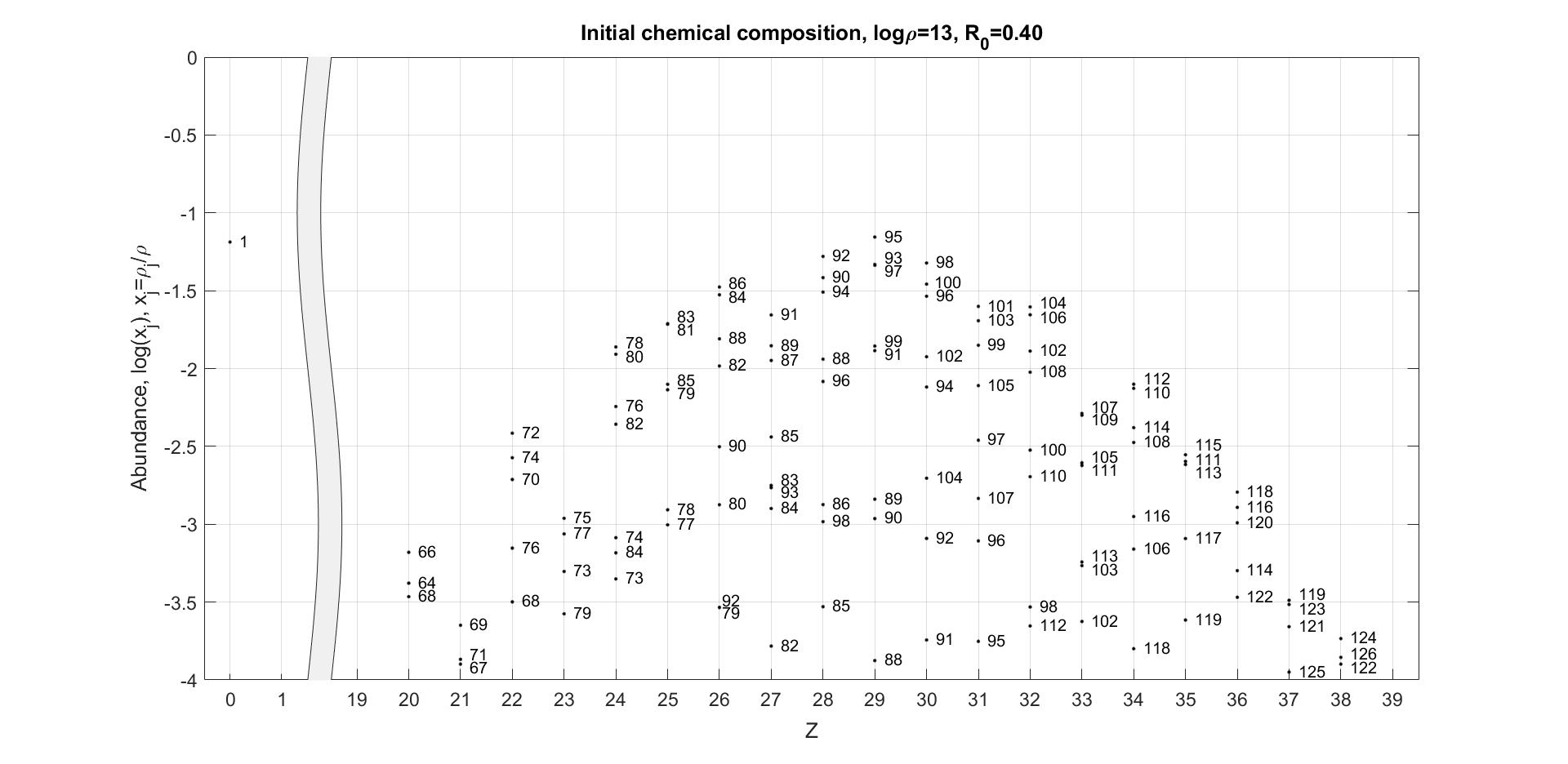}
   \caption{Initial chemical composition at $\mathrm{log \rho=13}$, $\mathrm{T_{9}=10}$ and $\mathrm{R_{0}=0.40}$. Each isotope is marked by the atomic mass number $\mathrm{A}$.}
  \label{Initial-10-13-40}
\end{figure*}

\begin{figure*}[hb!]
    \includegraphics[width=\textwidth,height=0.38\textheight]{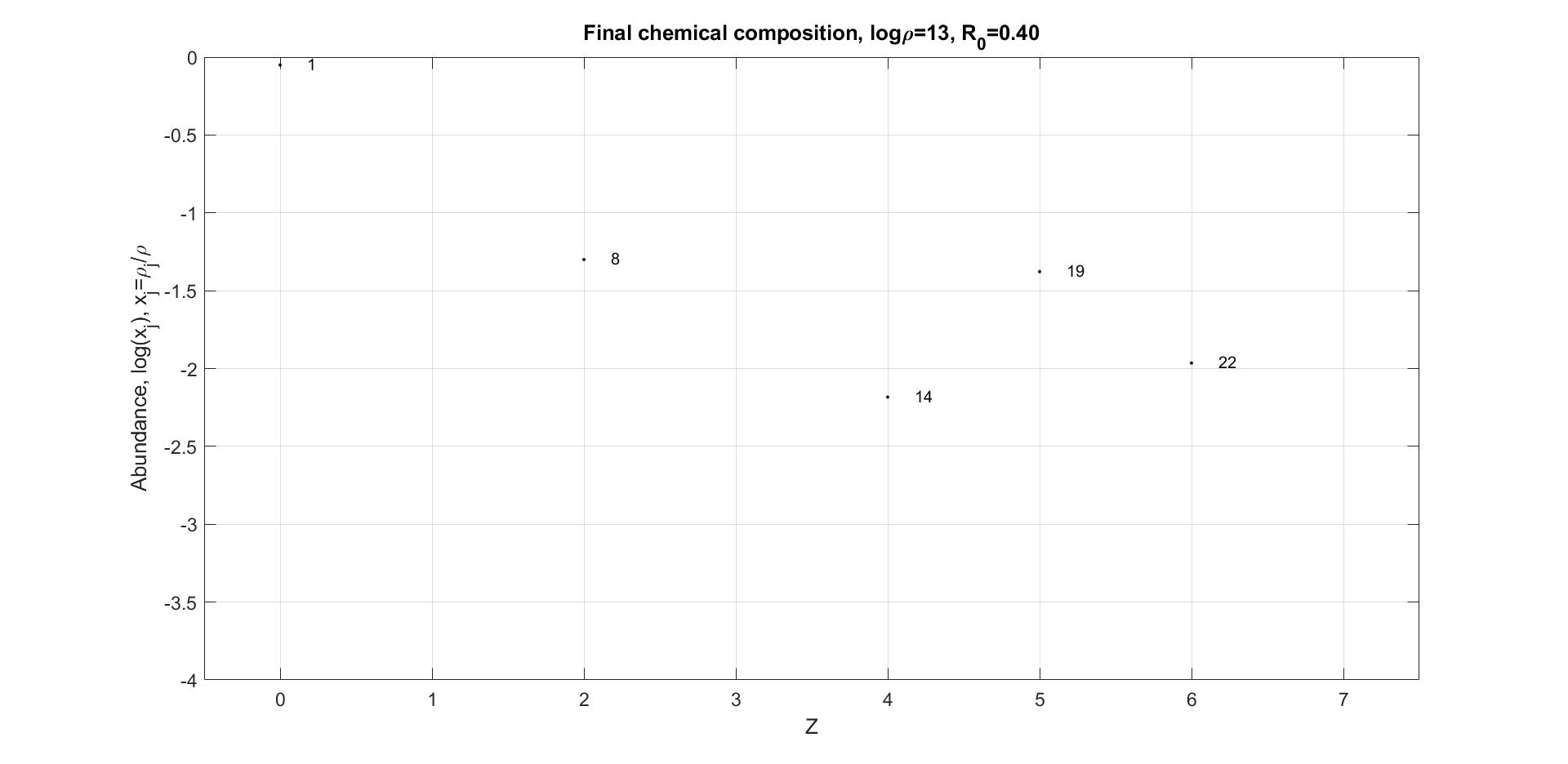}
   \caption{Chemical composition at the last calculated point $\mathrm {t = 200}$ s for $\mathrm{log \rho=13}$, $\mathrm{T_{9}=10}$ and $\mathrm{R_{0}=0.40}$. Each isotope is marked by the atomic mass number $\mathrm{A}$.}
  \label{Final-10-13-40}
\end{figure*}

Figures \ref{E_fem-10-11} and \ref{E_fem-10-13} correspond to the Fermi energy trajectories $\mathrm{\varepsilon_{fe}}$ for different parameter $\mathrm{R}$ and density $\mathrm{\rho}$ values. Figures \ref{Initial-10-11-05} - \ref{Final-10-11-33} and \ref{Initial-10-13-20} - \ref{Final-10-13-40} represents the initial and final chemical compositions of the shell layer for different parameter $\mathrm{R_{0}}$ initial values at the densities $\mathrm{log \rho=11}$ and $\mathrm{log \rho=13}$ respectively. All possible isotopes $\mathrm{Z \in [0;\;83]}$ that are in the equilibrium described by the Saha equation ($\ref{Saha}$) were chosen as seed nuclei. The density does not change during the calculations. The captions next to the labels in the figures mark the atomic mass numbers $\mathrm{A}$ of the isotopes.

As a result of calculations, a non-equilibrium composition of matter at a given density was obtained that is formed during cooling. The most prevalent nuclei have $\mathrm{A \geq 3Z}$ and are located next to the neutron drip line with low neutron binding energies $\mathrm{Q_{n} \approx 0}$, and $\mathrm{Q_{\beta} \approx \varepsilon_{fe}}$. The variety of resulting nuclei decreases with a temperature drop. The final composition of matter depends on the relative neutrons number of the initial composition.

In the case of a low density $\mathrm{log \rho=11}$, heavy nuclei are formed. The initial Fermi energies $\mathrm{\varepsilon_{fe}}$ (Fig. \Ref{E_fem-10-11}) are less than the beta decays energies of nuclei at the neutron drip line. As the temperature drops, the nuclei capture neutrons and, passing to the $\mathrm{Q_{n} \approx 0}$ boundary, undergo beta decay at $\mathrm{Q_{\beta}(A, Z)>\varepsilon_{fe}}$, increasing their charge atomic numbers $\mathrm{Z}$. The resulting nuclei also absorb neutrons, passing to the neutron drip line and decaying. The parameter $\mathrm{R}$ value increases. The Fermi energy increases (Fig. \Ref{R-10-11}, \ref{E_fem-10-11}) until it equals $\mathrm{Q_{\beta}(A, Z)}$. Beta decays stop at this point. The final composition strongly depends on the parameter $\mathrm{R_{0}}$ initial value, which determines the ratio of the nuclei number to free neutrons number at the transition from Saha equilibrium described by equation (\ref{Saha}) to quasi-equilibrium described by expression (\ref{SahaN}), when the total number of nuclei does not change anymore. The less parameter $\mathrm{R_{0}}$ value ($\mathrm{R_{0}=0.05}$ case, Fig. \ref{Initial-10-11-05} and \ref{Final-10-11-05}), the heavier nuclei are formed $\mathrm{A>230}$. The final chemical compositions for $\mathrm{R_{0}=0.20,\,0.33}$ (Fig. \ref{Final-10-11-20} and \ref{Final-10-11-33}) do not reach neutron drip line and match $\mathrm{A \approx 3Z}$, what refers to $\mathrm{R \approx 0.5}$ (Fig. \ref{R-10-11}). The reason is that there is insufficient amount of free neutrons remaining after the nuclear statistical equilibrium stage described by (\ref{Saha}). The lower free neutrons concentration $\mathrm{n_{n}}$, the fewer nuclei close to the neutron drip line are formed (\ref{SahaN}).

In contrast, at high density $\mathrm{log \rho=13}$ light elements are formed. The initial Fermi energy is $\mathrm{\varepsilon_{fe}>Q_{\beta}^{(max)}}$ and only electron captures are available to nuclei in degenerate matter. As a result, the nuclei decrease their charge atomic number $\mathrm{Z}$, passing to the neutron drip line $\mathrm{Q_{n} \approx 0}$ and emitting neutrons. The parameter $\mathrm{R}$ value decreases (Fig. \ref{R-10-13}) as well as $\mathrm{(A,Z)}$ until the reducing Fermi energy of electrons equals the threshold electron capture energies of light nuclei $\mathrm{\varepsilon_{fe} \approx \varepsilon_{\beta i}}$. As a result, the final nuclei distribution consists of free neutrons, helium $\mathrm{^{8} He}$, beryllium $\mathrm{^{14} Be}$, boron $\mathrm{^{19} B}$ and carbon $\mathrm{^{22}C}$ nuclei in different proportions that depend on the initial $\mathrm{R_{0}}$ (Fig. \ref{Final-10-13-20}, \ref{Final-10-13-33} and \ref{Final-10-13-40}). In our calculations, the neutrons have been considered in equilibrium with nuclei according to (\ref{SahaN}). So the final composition consists of several elements near the neutron drip line. A temperature drop below our limit of $\mathrm{T_{9}=0.1}$ leads to the situation when neutrons photodetachment can not occur. In this case, the final composition consists of only one element on the neutron drip line with $\mathrm{\varepsilon_{fe} \approx Q_{p0}}$ \cite{B-kogan1}. Maximum value of $\mathrm{Q_{p0}}$ in the used nucleus model corresponds to carbon $\mathrm{^{22}C}$ with $\mathrm{Q_{p}(22,6) \approx 32}$ MeV.

\vspace{0.1cm}

\center{6. CONCLUSION} \\ \justifying The analysis of the calculation results reveals that the problem of the chemical composition evolution of a hot neutron star crust at subnuclear densities, cooling due to neutrino energy loss, should be considered taking into account the previous history of the system. To obtain real initial conditions, it is necessary to have information on the dynamics of a collapsing supernova core and the parameters of a hot neutron star being born.

The properties of the non-equilibrium layer obtained within the framework of this model indicate the presence of a large reserve of neutron stars nuclear energy, capable of maintaining the X-ray luminosity for a long period of time.

As a result of various stellar activity processes, such as starquakes, elements from the shell can be carried outward, leading to the appearance of explosive transient processes, as well as forming heavy elements, such as transuranium elements, that are part of the next generation objects, including the Earth.
\center{ACKNOWLEDGEMENTS} \\ \justifying The authors are grateful to I. V. Panov for providing the nuclear data used in this article. The work of G. S. Bisnovatyi-Kogan was carried out with the partial support of the RFBR grant 20-02-00455, the work of A. Yu. Ignatovskiy was supported by the RSF grant 21-12-00061.
\selectlanguage{russian}

\end{multicols}

\end{document}